\documentclass[draft]{agujournal2019}

\usepackage{url} 
\usepackage{lineno}
\usepackage[inline]{trackchanges} 
\usepackage{soul}
\usepackage{longtable}

\draftfalse

\journalname{Space Weather}

\begin{document}

\title{Comparing 1-year GUMICS$-$4 simulations of the Terrestrial Magnetosphere with Cluster Measurements}

\authors{G.~Facsk{\'o}\affil{1,2,3,4}, D.~G.~Sibeck\affil{3}, I.~Honkonen\affil{5}, J.~B{\'o}r\affil{6}, G.~Farinas~Perez\affil{3,4}\thanks{Now at University of Miami, Electrical and Computer Engineering Department, Miami, Florida, USA}, A.~Tim{\'a}r\affil{1}, Y.~Y.~Shprits\affil{7,8}, P.~Peitso\affil{9}\thanks{Now at Aurora Propulsion Technologies, Espoo, Finland}, L.~Degener\affil{5}\thanks{Now private individual, Hannover, Germany}, E.~I.~Tanskanen\affil{10}, C.~R.~Anekallu\affil{11}, S.~Szalai\affil{6,12}, {\'A.}~Kis\affil{6}, V.~Wesztergom\affil{6}, {\'A}.~Mad{\'a}r\affil{1,13}, N.~Bir{\'o}\affil{1,13}, G.~Kob{\'a}n\affil{1,13}, A.~Illy{\'e}s\affil{1}, P.~Kov{\'a}cs\affil{1}, Z.~D{\'a}lya\affil{1}, M.~Lkhagvadorj\affil{1,14}} 

\affiliation{1}{Wigner Research Centre for Physics, Budapest, Hungary}
\affiliation{2}{Milton Friedman University, Budapest, Hungary}
\affiliation{3}{NASA Goddard Space Flight Center, Greenbelt, Maryland, USA}
\affiliation{4}{the Catholic University of America, Washington DC, USA}
\affiliation{5}{Finnish Meteorological Institute, Helsinki, Finland}
\affiliation{6}{Institute of Earth Physics and Space Science (ELKH EPSS), Sopron, Hungary}
\affiliation{7}{Helmholtz Centre Potsdam, GFZ German Research Centre for Geosciences, Potsdam, Germany}
\affiliation{8}{Institute for Physics and Astronomy, University of Potsdam, Potsdam, Germany}
\affiliation{9}{Aalto University, School of Electrical Engineering, Espoo, Finland}
\affiliation{10}{Sodankyl{\"a} Geophysical Observatory, University of Oulu, Sodankyl{\"a}, Finland}
\affiliation{11}{UCL Department of Space \& Climate Physics, Mullard Space Science Laboratory, Dorking, UK}
\affiliation{12}{University of Miskolc, Department of Geophysics, Miskolc, Hungary}
\affiliation{13}{E{\"o}tv{\"o}s Lor{\'a}nd University, Doctoral School of Physics, Budapest, Hungary}
\affiliation{14}{E{\"o}tv{\"o}s Lor{\'a}nd University, Institute of Physics, Budapest, Hungary}

\correspondingauthor{G{\'a}bor Facsk{\'o}}{facsko.gabor@wigner.hu}

\begin{keypoints}
\item The GUMICS$-$4 code provides realistic ion plasma moments and magnetic field in the solar wind and the outer magnetosheath.
\item The code predicts realistic bow shock locations.
\item An inner magnetosphere model should be added to the code to increase the accuracy of the simulation in the inner magnetosphere.
\end{keypoints}

\begin{abstract}
We compare the predictions of the GUMICS$-$4 global magnetohydrodynamic model for the interaction of the solar wind with the Earth's magnetosphere with Cluster~SC3 measurements for over one year, from January 29, 2002, to February 2, 2003. In particular, we compare model predictions with the north/south component of the magnetic field ($B_{z}$) seen by the magnetometer, the component of the velocity along the Sun-Earth line ($V_{x}$), and the plasma density as determined from a top hat plasma spectrometer and the spacecraft's potential from the electric field instrument. We select intervals in the solar wind, the magnetosheath, and the magnetosphere where these instruments provided good quality data and the model correctly predicted the region in which the spacecraft is located. We determine the location of the bow shock, the magnetopause and, the neutral sheet from the spacecraft measurements and compare these locations to those predicted by the simulation.

The GUMICS$-$4 model agrees well with the measurements in the solar wind however its accuracy is worse in the magnetosheath. The simulation results are not realistic in the magnetosphere. The bow shock location is predicted well, however, the magnetopause location is less accurate. The neutral sheet positions are located quite accurately thanks to the special solar wind conditions when the $B_{y}$ component of the interplanetary magnetic field is small. 
\end{abstract}

\section*{Plain Language Summary}

We compare output from a model for the Earth's space environment with the spacecraft observations of the magnetic field strength and direction, solar wind velocity, and two different density measurements over the course of 1 year. We select intervals from locations in regions near Earth where the spacecraft instruments provide high quality data and the model correctly predict the region in which the spacecraft is located. We identify the locations where the spacecraft observes boundaries between different regions and compare these locations to those predicted by the simulation. The model agrees well with the measurements in the solar wind, but its accuracy diminishes in the slower, thermalized, and compressed flow around the region dominated by the Earth's magnetic field. In this region, the model does not seem to be realistic. The locations of the boundaries are generally good, but predictions for the location of the boundary of the region dominated by the terrestrial magnetic field and the domain of the slower, compressed solar wind stream are less accurate. 

\section{Introduction}
\label{sec:introduction}

One of the most cost-effective ways to study the interaction of the solar wind with planetary magnetospheres (or predict conditions in near-Earth space) is modeling this complex system using a magnetohydrodynamic (MHD) code. In the past, several parallelized codes were developed, which are used for forecasting the near-Earth space environment. Such as the Lyon-Fedder-Mobarry \cite<LFM;>[]{lyon04:_lyon_fedder_mobar_lfm_mhd} code, the Grid Agnostic MHD for Extended Research Applications \cite<GAMERA;>[]{zhang19:_gamer}, the Open Geospace General Circulation Model \cite<OpenGGCM;>[]{raeder08:_openg_simul_themis_mission}, or the Block-Adaptive-Tree-Solarwind-Roe-Upwind-Scheme \cite<BATS-R-US;>[]{powell99:_solut_adapt_upwin_schem_ideal_magnet,toth12:_adapt}. In Europe three global MHD codes have been developed: the Grand Unified Magnetosphere--Ionosphere Coupling Simulation \cite<GUMICS$-$4;>[]{janhunen12:_gumic_mhd}, the Computational Object--Oriented Libraries for Fluid Dynamics \cite<COOLFluiD;>[]{lani12:_coolf_open_comput_platf_aerot} and the 3D resistive magnetohydrodynamic code Gorgon \cite{chittenden04:_x_z,ciardi07}. The COOLFluiD is a general-purpose plasma simulation tool. The Gorgon code was developed to study high--energy, collisional plasma interactions and has been adapted to simulate planetary magnetospheres and their interaction with the solar wind \cite{mejnertsen16:_global_mhd_neptun,mejnertsen18:_global_mhd_simul_earth_bow}. Neither Gorgon nor COOLfluid has an ionospheric solver. Almost all of these codes are available at the Community Coordinated Modelling Center (CCMC; http://ccmc.gsfc.nasa.gov/) hosted by the NASA Goddard Space Flight Center (GSFC) or the Virtual Space Weather Modelling Centre (VSWMC; http://swe.ssa.esa.int/web/guest/kul-cmpa-federated; requires registration for the European Space Agency (ESA) Space Situational Awareness (SSA) Space Weather (SWE) portal) hosted by the KU Leuven \cite{poedts20:_virtual_space_weath_model_centr}. A comparison of the simulation results with spacecraft and ground-based measurements is necessary to understand the abilities and features of the developed tools. A statistical study using long--term global MHD runs for validation of the codes seems necessary. Because providing long simulations is costly and time--consuming, only a few studies have been done, almost all for periods much less than a year except \citeA{liemohn18:_real_time_swmf_ccmc}.

\citeA{guild08:_geotail_lfm1,guild08:_geotail_lfm2} launched two months of LFM runs and compared the plasma sheet properties in the simulated tail with the statistical properties of six years of Geotail magnetic field and plasma observations \cite{kokubun94:_geotail_magnet_field_exper,mukai94:_low_energ_partic_lep_exper_geotail_satel}. The LFM successfully reproduced the global features of the global plasma sheet in a statistical sense. However, there were some differences. The predicted plasma sheet was too cold, too dense, and the bulk flow was faster than the observed plasma sheet \cite{kokubun94:_geotail_magnet_field_exper,mukai94:_low_energ_partic_lep_exper_geotail_satel}. The LFM overestimated the ionospheric transpolar potential. The transpolar potential correlated with the speed of the plasma sheet flows. Equatorial maps of density, thermal pressure, thermal energy and, velocity were compared. The LFM overestimated the plasma sheet density close to the Earth, the temperature by a factor of $\sim$3 and the global average flow speed by a factor of $\sim$2. The LFM reproduced many of the climatological features of the Geotail data set. The low-resolution model underestimated the occurrence of the fast earthward and tailward flows. Increasing the simulation resolution resulted in the development of fast, bursty flows. These flows influenced the statistics and contributed to a better agreement between simulations and observations.

\citeA{zhang11:_lyon_fedder_mobar_mhd} studied the statistics of magnetosphere-ionosphere (MI) coupling using the LFM simulation of \citeA{guild08:_geotail_lfm1} above. The polar cap potential and the field--aligned currents (FAC), the downward Poynting flux and, the vorticity of the ionospheric convection were compared with observed statistical averages and the Weimer05 empirical model \cite{weimer05:_improv_joule}. The comparisons showed that the LFM model produced quite accurate average distributions of the Region 1 (R1) and Region 2 (R2) currents. The ionospheric R2 currents in the MHD simulation seemed to originate from the diamagnetic ring current. The average LFM R1 and R2 currents were small compared with the values from the Weimer05 model. The average Cross Polar Cap Potential (CPCP) was higher in the LFM simulation than the measurements of the SuperDARN and the Weimer05 model. The average convention pattern was quite symmetric in the LFM simulation as compared to the SuperDARN measurements and the Weimer05 model. The SuperDARN measurements and the Weimer05 model had a dawn-dusk asymmetry. In the LFM model, more Poynting flux flowed into the polar region ionosphere than in the Weimer05 model as a consequence of the larger CPCP in the LFM simulation. The larger CPCP allowed a higher electric field in the polar region. The statistical dependence of the high-latitude convection patterns on Interplanetary Magnetic Field (IMF) clock angle was similar to the SuperDARN measurements \cite{sofko95:_direc_super} and the Weimer05 model. The average ionospheric field-aligned vorticity showed good agreement on the dayside. However, the LFM model gave larger nightside vorticity than SuperDARN measurements because the Pedersen conductance on the night side ionosphere was too low. 

\citeA{wiltberger17:_struc_high_latit_curren_magnet_ionos_model} studied the structure of high latitude field-aligned current patterns using three resolutions of the LFM global MHD code and the Weimer05 empirical model \cite{weimer05:_improv_joule}. The studied period was a month$-$long and contained two high-speed streams. Generally, the patterns agreed well with results obtained from the Weiner05 computing. As the resolution of the simulations increased, the currents became more intense and narrow. The ratio of the Region 1 (R1), the Region 2 (R2) currents and, the R1/R2 ratio increased when the simulation resolution increases. However, both the R1 and R2 currents were smaller than the predictions of the Weimer05 model. This effect led to a better agreement of the LFM simulation results with the Weimer~2005 model results. The CPCP pattern became concentrated in higher latitudes because of the stronger R2 currents. The relationship of the CPCP and the R1 looked evident at a higher resolution of the simulation. The LFM simulation could have reproduced the statistical features of the field$-$aligned current (FAC) patterns. 

\citeA{haiducek17:_swmf_global_magnet_simul_januar} simulated the month of January 2005 using the Space Weather Modelling Framework \cite<SWMF;>[]{toth05:_space_weath_model_framew} and the OMNI solar wind data (https://omniweb.gsfc.nasa.gov/) as input. The simulations were executed with and without an inner magnetosphere model and using two different grid resolutions in the magnetosphere. The model was very good in predicting the ring currents \cite<SYM-H; http://wdc.kugi.kyoto-u.ac.jp/aeasy/asy.pdf;>[]{iyemori90:_storm}. The $K_p$ index (a measure of the general magnetospheric convention and the auroral currents \cite{bartels39,rostoker72:_geomag,thomsen04:_why_kp}) was predicted well during storms however the index was overestimated during quiet periods. The AL index (that describes the westward electrojet of the surface magnetic field introduced by \citeA{davis66:_auror_ae}) was predicted reasonably well on average. However, the model reached the highest negative AL value less often than it was reached in observations because the model captured the structure of the auroral zone currents poorly. The overpredicting of $K_p$ index during quiet times might have happened for the same reason because it is also sensitive to auroral zone dynamics. The SWMF usually over$-$predicted the CPCP. These results were not sensitive to grid resolutions, except for of the AL index, which reached the highest negative value more often when the grid resolution was higher. Switching the inner magnetosphere model off had a negative effect on the accuracy of all quantities mentioned above, except the CPCP. 

This paper compares the Cluster SC3 measurements directly to a previously made 1-year long GUMICS$-$4 simulation at locations in the solar wind, magnetosheath, and the magnetosphere along the Cluster SC3 orbit \cite{facsko16:_one_earth}. The parameters are $B_z$, the north/south component of the magnetic field in GSE coordinates, the solar wind velocity GSE X component ($V_x$), and the solar wind density $n$. We also compare the predicted and observed locations of the bow shock, magnetopause, and the neutral sheet. These parameters are selected because $B_z$ controls the solar wind--magnetosphere interaction, $V_x$ is the main component of the solar wind velocity and $n$ is the ion plasma moment that is the easiest to calculate; furthermore, several instruments could determine it (see Section~\ref{sec:cluster}). The structure of this paper is as follows. Section~\ref{sec:data} describes the GUMICS$-$4 code, the 1-year simulation, and the Cluster spacecraft measurements. Section~\ref{sec:comp} gives comparisons between the simulations and observations. Results of the comparison are discussed in Section~\ref{sec:discussion}. Finally, Section~\ref{sec:concl} contains the conclusions.

\section{The GUMICS$-$4 products and Cluster measurements}
\label{sec:data}

Here we use two very different time series. The first type is derived from a previous 1-year run of the GUMICS$-$4 simulation \cite{facsko16:_one_earth}. The second time series was measured by the magnetometer, ion plasma, and electric field instruments of the Cluster reference spacecraft.

\subsection{The GUMICS$-$4 code}
\label{sec:gumics}

The GUMICS$-$4 model has two coupled simulation domains, the magnetospheric domain outside of a 3.7\,$R_E$ radius sphere around the Earth, and a coupled ionosphere module containing a 2D height-integrated model of ionosphere. GUMICS$-$4 is not a parallel code model however it has been extensively used to study energy propagation from the solar wind into the magnetosphere through the magnetopause and other features \cite[see the references therein]{janhunen12:_gumic_mhd}. The code has also been applied to study forced reconnection in the tail \cite{voeroes14:_winds_condit_ram_co_ram}. Recently, several hundred synthetic two hours duration GUMICS$-$4 simulation runs were made to compare the simulation results to empirical formulas \cite{gordeev13:_verif_gumic_mhd}. The agreement was quite good in general, but the diameter of the magnetopause in the simulations deviated slightly (10\,\%) from corresponding observations in the tail. The GUMICS$-$4 simulation magnetotail was smaller than that which the spacecraft observed. However, the modeled magnetopause showed good agreement with the empirical model in the mid--tail at northward IMF conditions. \citeA{facsko16:_one_earth} made a 1-year long simulation using the GUMICS$-$4 code. In those simulations, the magnetotail was significantly shorter than that which the spacecraft observed \cite{facsko16:_one_earth}. \citeA{gordeev13:_verif_gumic_mhd} and \citeA{voeroes14:_winds_condit_ram_co_ram} had similar experience when the simulations of the papers were evaluated. \citeA{juusola14:_statis_gumic_mhd} compared the ionospheric currents, fields and the Cross Polar Cap Potential Drop (CPCP) in the simulation to observations from the Super Dual Auroral Radar Network (SuperDARN) radars \cite{greenwald95:_darn_super} and CHAMP spacecraft \cite{reigber02:_champ} observations of field--aligned currents (FAC) \cite{juusola07:_hall_peder_champ,ritter04:_ionos_champ_image}. The CPCP, the FAC, and other currents could not be reproduced properly. A possible cause for this poor agreement could be the model's low resolution in the inner magnetosphere and/or the lack of an inner magnetosphere model accurately incorporating the physics of this region. This hypothesis is supported by the result of \citeA{haiducek17:_swmf_global_magnet_simul_januar}. \citeA{haiducek17:_swmf_global_magnet_simul_januar} simulated only a month--long period using a different spatial resolution and tested the code with the inner magnetosphere model of the SWMF switched off for a special run. This run without an inner magnetosphere model made it clear that only the CPCP parameter of the simulation agreed quite well with the measurements. This fact explained why the agreement between the Cluster SC3 and the GUMICS-4 simulations was so good as described by \citeA{lakka18:_cross_polar_cap_satur_gumic,lakka18:_icme_earth_mach} based on the CPCP in GUMICS$-$4 simulations. \citeA{kallio15:_proper} determined plasma and magnetic field parameters along the lunar orbit from \citeA{facsko16:_one_earth}'s global MHD simulations. The parameters differed significantly from observations in the magnetotail indicating the need for future studies. \citeA{facsko16:_one_earth} determined the footprint of Cluster SC3 using the 1-year simulation and the Tsyganenko T96 empirical model \cite{tsyganenko95:_model_earth}. The agreement of the footprint was better in the Northern Hemisphere. The GUMICS$-$4 tail was shorter in the simulations than the observations. 

A 1-year global MHD simulation was produced with the GUMICS$-$4 code using the OMNI solar wind data from January 29, 2002, to February 2, 2003, as input \cite{facsko16:_one_earth}. The creation and analysis of the simulation were based on a work package of the European Cluster Assimilation Technology (ECLAT) project (https://cordis.europa.eu/result/rcn/165813\_en.html; http://www.eclat--project.eu/). The GUMICS-4 is a serial code \cite{janhunen12:_gumic_mhd} hence the 1-year simulation was made in 1860 independent runs. This interval covered 155 Cluster SC3 orbits and each orbit lasted 57 hours. The FMI supercomputer at the time had 12 cores on each node hence the 57 hours were divided into 4.7 hours simulation time with one hour initialization period. Each sub-interval used its own individual average Geocentric Solar Ecliptic (GSE) IMF magnetic field X component $B_x$ component and dipole tilt angle. All data gaps in the solar wind were interpolated linearly. If the data gap of the input file was at the beginning (or the end) of the interval then the first (or last) good data from the input file was used to fill the gap. The initialization of each simulation run was made using constant values. These values were the first valid data of the input file repeated 60 times (60 minutes) in the input file of the sub-interval. The simulation results were saved every five minutes. Various simulation parameters, for example, the density, particle density, temperature, magnetic field, solar wind velocity (29 different quantities) were saved from the simulation results along the Cluster reference spacecraft's orbit in the GSE coordinates. 

\subsection{The Cluster SC3 measurements}
\label{sec:cluster}

The Cluster-II mission of the European Space Agency (ESA) was launched in 2000 to observe geospace \cite{credland97:_clust_mission,escoubet01:_introd_clust}. The four spacecraft form a tetrahedron in space however here we use only the measurements of the reference spacecraft, Cluster SC3. The spacecraft was spin--stabilized and its rotation period is $\sim$4\,s. Hence, the intrinsic time resolution of the plasma instruments is 4\,s and we use 4\,s averaged magnetic field data. The highest resolution of the Cluster FluxGate Magnetometer (FGM) magnetic field instrument is 27\,Hz \cite{balogh97:_clust_magnet_field_inves,balogh01:_clust_magnet_field_inves}. The ion plasma data are provided by the Cluster Ion Spectrometry (CIS) Hot Ion Analyser (HIA) sub-instrument \cite{reme97:_clust_ion_spect_exper,reme01:_first_earth_clust_cis}. The CIS HIA instrument is calibrated using the Waves of HIgh frequency Sounder for Probing the Electron density by Relaxation (WHISPER) wave instrument onboard Cluster \cite{decreau01:_early_whisp_clust,trotignon10:_whisp_relax_sound_clust_activ_archiv,blagau13:_in_hot_ion_analy_clust,blagau14:_in_hot_ion_analy_clust}. The results of these calibrations can appear as sudden non-physical jumps in the CIS HIA data. The CIS HIA had different modes to measure in the solar wind and the magnetosphere. When the instrument is switched from one mode to another mode non-physical jumps also appear in the measurements. These features impair the accuracy of data analyses.

We remove non-physical jumps from our results using a density determination based on different principles. We use the spacecraft potential of the Electric Field and Wave Experiment \cite<EFW ;>[]{gustafsson97:_elect_field_wave_exper_clust_mission,gustafsson01:_first_clust_efw} to determine the electron density. This quantity can be calculated using the empirical density formula 
\begin{equation}\label{eq:empdens}
n_{EFW}=200(V_{sc})^{-1.85},
\end{equation}
where $n_{EFW}$ is the calculated density and $V_{sc}$ is the Cluster EFW spacecraft potential \cite{trotignon10:_whisp_relax_sound_clust_activ_archiv,trotignon11:_calib_repor_whisp_measur_clust}. The EFW and the WHISPER were used for the calibration of the CIS HIA and the Plasma Electron and Current Experiment \cite<PEACE;>[]{johnstone97:_peace,fazakerley10:_peace_data_clust_activ_archiv,fazakerley10:_clust_peace_in_calib_status}. Both instruments were still working onboard all Cluster spacecraft. Their stable operation reduced the number of data gaps, and it also made the data analysis easier.

\section{Comparison of measurements to simulation}
\label{sec:comp}

The parameters saved from the GUMICS$-$4 simulations and the Cluster SC3 magnetic field, solar wind velocity and, density measurements are compared in different regions, namely the solar wind, magnetosheath, and magnetosphere via cross$-$correlation calculations. The temporal resolution of the simulated Cluster orbit data is mostly five minutes because the results of the simulations are saved every five minutes \cite{facsko16:_one_earth}. However, the time difference between points can be more than five minutes at the boundary of the subintervals, because the length of the simulation intervals is determined in minutes. To facilitate analysis of the simulation results, all simulation data were interpolated to a one$-$minute resolution. This method does not provide extra information to the cross$-$correlation calculation. The data gaps are eliminated using linear interpolation and extrapolation when the gap is at the start or the end of the selected interval. The spin resolution (4\,s) of Cluster SC3 magnetic field measurements is averaged over five minutes around ($\pm150\,s$) the timestamps of the saved data. Then the averaged data were interpolated to a one$-$minute resolution to make the correlation calculations.

For the correlation calculation, intervals are selected carefully in the solar wind (see Section~\ref{sec:sw}), the magnetosheath (see Section~\ref{sec:msh}), the dayside and the night side magnetosphere (see Section~\ref{sec:msph}). In these intervals, the parameters did not vary a lot and we exclude intervals when either Cluster or the virtual probe cross any boundaries. To compare the $B_z$ magnetic field, $V_x$ solar wind speed and the $n_{CIS}$ and the $n_{EFW}$ curves we cross--correlate selected intervals. We carefully examine such cases, and remove intervals that are shorter than four hours for the $\pm$60\,minutes correlation calculation, and intervals with large data gaps from the correlation calculation. Those intervals are also excluded when the plasma instrument has a calibration error or a change in its recording mode as it moves from the magnetosphere to solar wind (for example). The electron density is also calculated using Equation~\ref{eq:empdens} and correlated. We want to avoid the calibration errors and sudden non-physical jumps mentioned previously. The correlation results for the density derived from the electric field potential results do not differ significantly from those for the top hat plasma instrument, however, the EFW's $n_{EFW}$ experiences no mode changes and it is applicable in the magnetosphere too (in contrast to the CIS HIA instrument).

\subsection{Solar wind}
\label{sec:sw}

\begin{figure}[h]
\centering
\includegraphics[width=0.9\textwidth,angle=0]{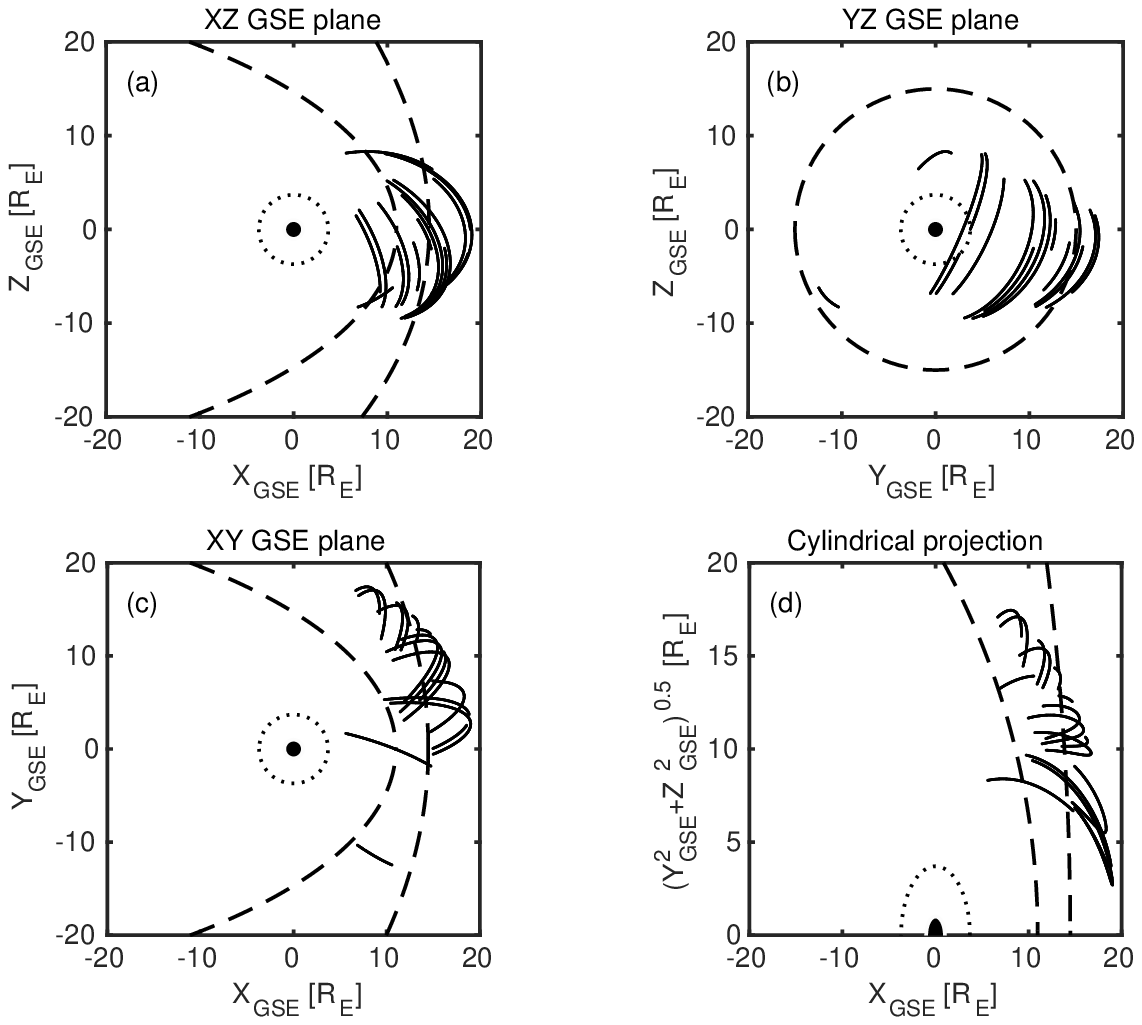} 
\caption{Cluster SC3 orbit in the solar wind in GSE system for all intervals (see Table~\ref{tab:sw}). (a) XZ (b) YZ (c) XY (d) Cylindrical projection. Average bow-shock and magnetopause positions are drawn on all plots using dashed lines \cite[respectively]{peredo95:_three_alfven_mach,tsyganenko95:_model_earth}. The black dots at $3.7\,R_E$ show the boundary of the GUMICS$-$4 inner magnetospheric domain. The black circle in the origo of all plots shows the size of the Earth.}
\label{fig:sworbit}
\end{figure}

We use OMNI IMF and solar wind velocity, density, and temperature data as input to the simulation. Comparing parameters obtained from the simulation and the measurements in the solar wind region is especially interesting because the IMF X component cannot be given to the GUMICS$-$4 as input \cite{janhunen12:_gumic_mhd,facsko16:_one_earth}. However, the magnetic field of the solar wind has an X component in the simulations. Additionally, solar wind structures might evolve from the simulation domain boundary at +32\,$R_E$ to the sub-solar point of the terrestrial bow shock where all OMNI data is shifted. Almost the same solar wind time intervals are used as in Table~1 of \citeA{facsko16:_one_earth}. Although the Cluster instruments were calibrated in 2002, just after launch, there are not many CIS HIA moment observations in 2001 and 2002 (Table~\ref{tab:sw}). Hence, we do not have satisfactory ion plasma data coverage for this year. Additionally, to improve the accuracy of the correlation calculation (see below) we omitted intervals (shorter than five hours) and those in which the CIS HIA instrument changed its mode. The Cluster fleet is located in the solar wind only from December to May and only for a couple of hours during each orbit near apogee. We double$-$check whether Cluster SC3 remains in the solar wind in both the simulation and reality. We also check the omnidirectional CIS HIA ion spectra on the Cluster Science Archive (CSA; https://www.cosmos.esa.int/web/csa/csds-quicklook-plots). The spectra must contain one narrow band in the solar wind region, indicating an observation of the solar wind beam. Hence, there are only 17 solar wind intervals to study, as shown in Figure~\ref{fig:sworbit}. 

\begin{figure}[h]
\centering
\includegraphics[width=0.9\textwidth,angle=0]{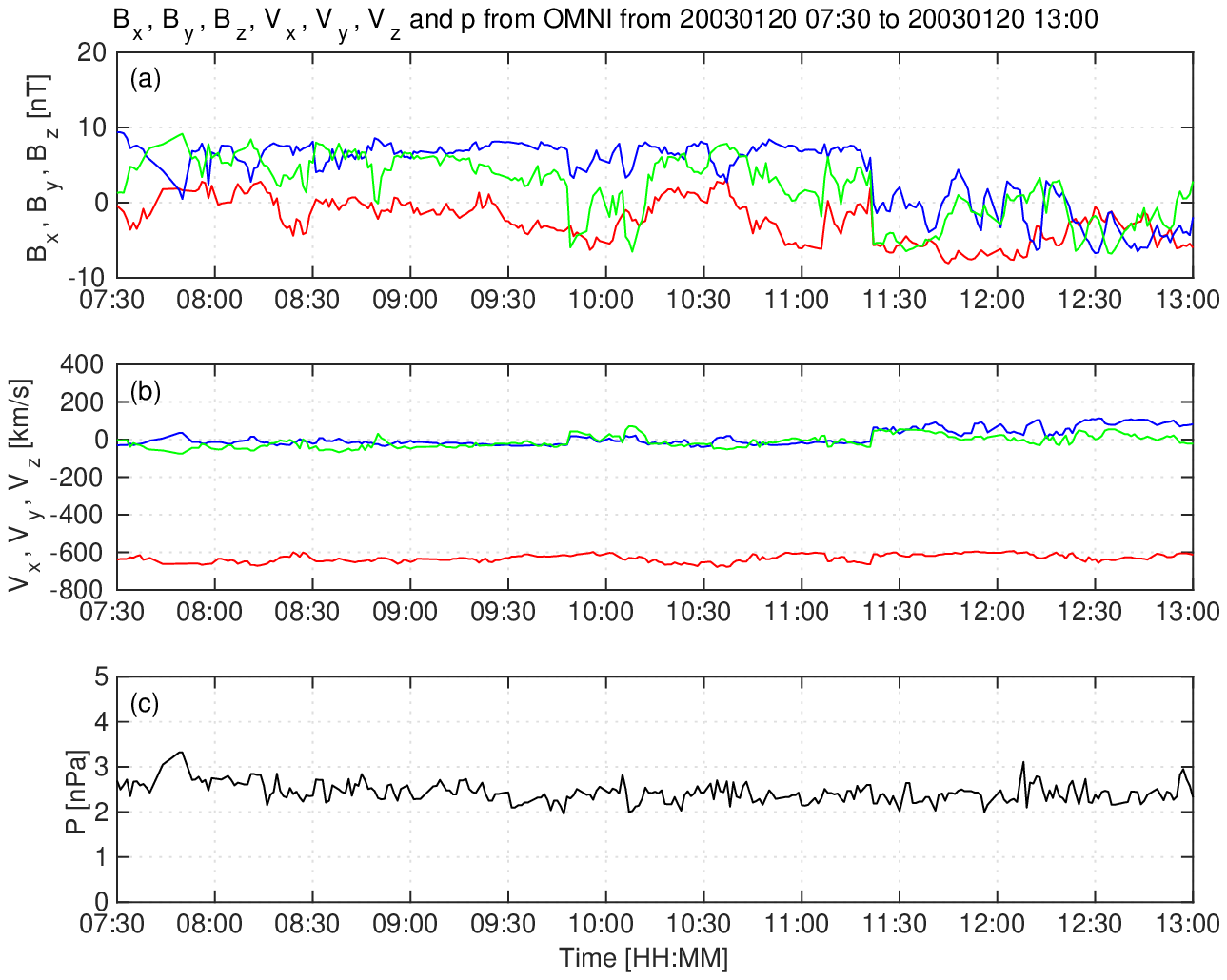} 
\caption{OMNI solar wind data in GSE system from 7:30 to 13:00 (UT) on January 20, 2003. (a) Magnetic field $B_{x}$ (red), $B_{y}$ (green) and $B_{z}$ (blue) components. (b) Solar wind velocity $V_{x}$ (red), $V_{y}$ (green) and $V_{z}$ (blue) components. (c) The $P$ pressure of the solar wind (black).}

\label{fig:swomni}
\end{figure}

\begin{figure}[h]
\centering
\includegraphics[width=0.9\textwidth,angle=0]{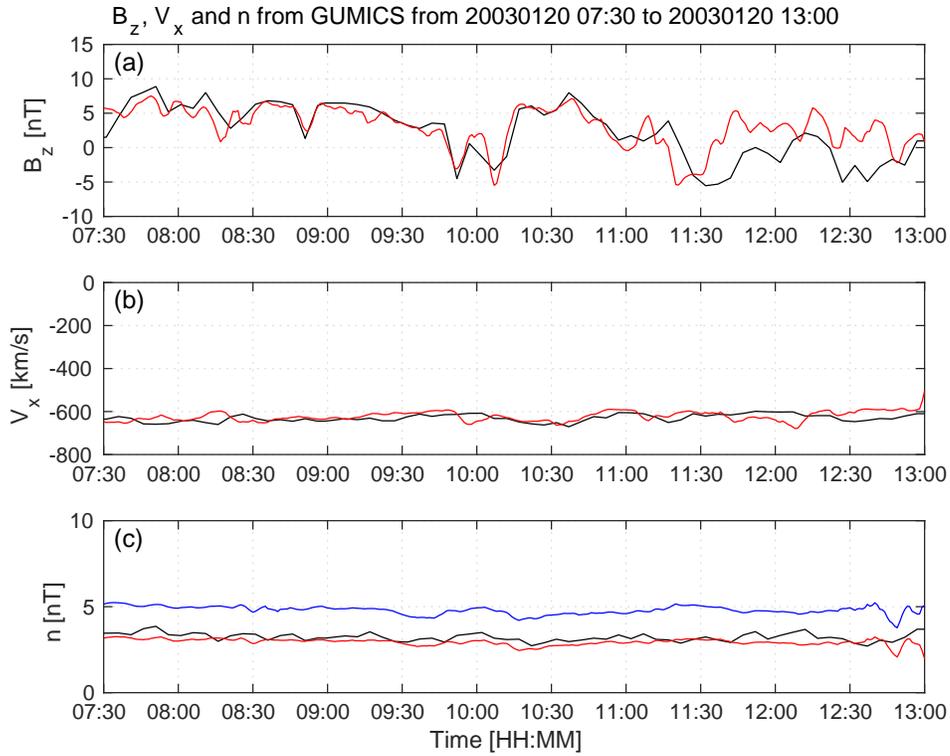} 
\caption{GUMICS-4 simulation results (black) and Cluster SC3 magnetic field Z component, ion plasma moments (red) and electron density calculated from spacecraft potential (blue) from January 20, 2003 from 7:30 to 13:00 (UT) in the solar wind in GSE system. (a) Magnetic field Z component. (b) Solar wind velocity X component (c) Solar wind density.}
\label{fig:swplot}
\end{figure}

\begin{figure}[h]
\centering
\includegraphics[width=0.9\textwidth,angle=0]{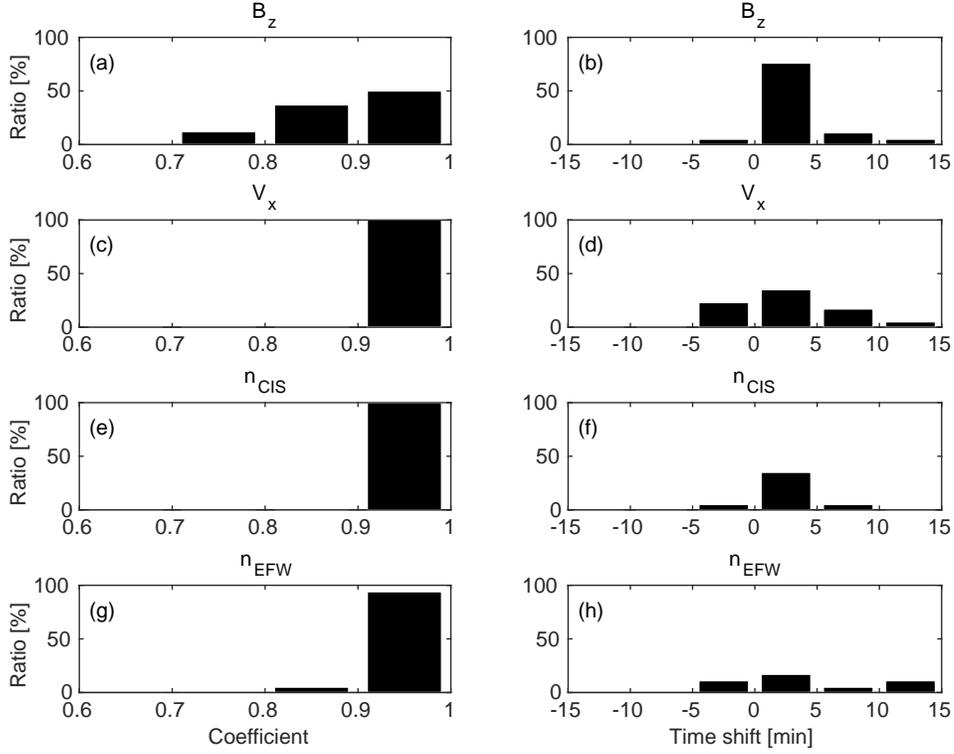} 
\caption{The distributions of the highest cross-correlation coefficients (a,~c,~e,~g) of the magnetic field Z component ($B_z$) in GSE system, solar wind velocity X component ($V_X$) in GSE system, the solar wind density measured by the CIS HIA ($n_{CIS}$) instrument and calculated from the spacecraft potential ($n_{EFW}$), respectively, for all intervals in the solar wind. The distributions of the corresponding time shifts (b,~d,~f,~h) of the $B_z$, the $V_X$, the $n_{CIS}$ and the $n_{EFW}$), respectively, for all intervals in the solar wind.}
\label{fig:swcorrplot}
\end{figure}

The selected intervals occur for quiet solar wind conditions (Figure~\ref{fig:swomni}). The GUMICS$-$4 simulation results have five--minute time resolution and the Cluster SC3 measurements have one--minute time resolution (Figure~\ref{fig:swplot}). The measurements vary significantly. Despite the quiet conditions the observed solar wind density often changes and deviates from the simulation. Figure~\ref{fig:swscatplot}c shows that both densities deviate significantly. The CIS HIA density variations are even larger as expected given the complexity and a large number of working modes of the CIS instrument. The magnetic field and the solar wind velocity fit better. Figure~\ref{fig:swcorrplot}a shows that the correlation of the magnetic fields is very good; furthermore on Figure~\ref{fig:swcorrplot}c,~\ref{fig:swcorrplot}e,~\ref{fig:swcorrplot}g the correlation of the solar wind velocity and density is excellent (Table~\ref{tab:sw}). The time shift in Figure~\ref{fig:swcorrplot}b, Figure~\ref{fig:swcorrplot}d, Figure~\ref{fig:swcorrplot}f is about five--minutes for the magnetic field and the CIS data. In Figure~\ref{fig:swcorrplot}h for the EFW data, the time--shift is less stable. It is not as well determined as in the case of the other parameters.

\subsection{Magnetosheath}
\label{sec:msh}

\begin{figure}[h]
\centering
\includegraphics[width=0.9\textwidth,angle=0]{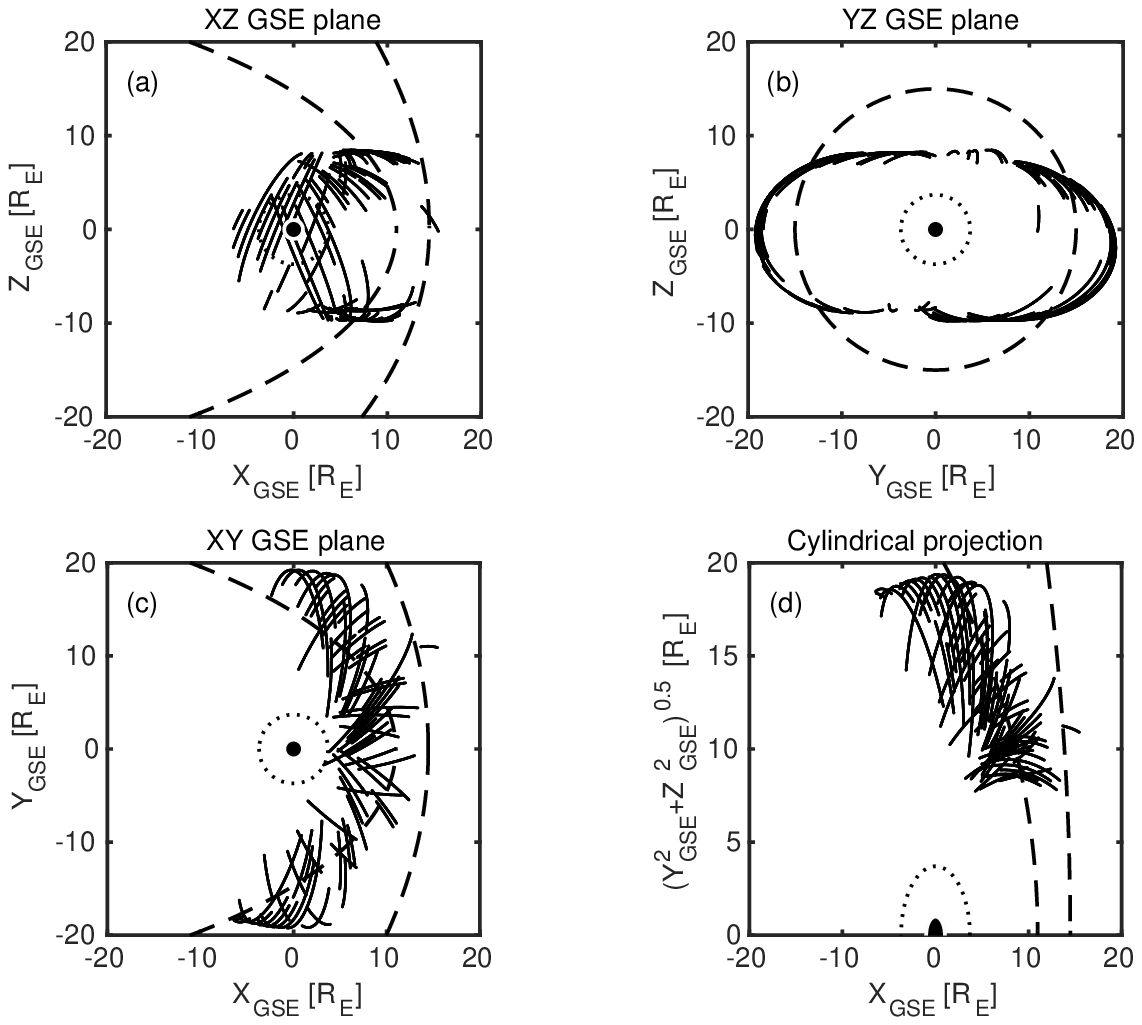} 
\caption{Cluster SC3 orbit in the magnetosheath in GSE system for all intervals (see Table~\ref{tab:msh}). (a) XZ (b) YZ (c) XY (d) Cylindrical projection. Average bow-shock and magnetopause positions are drawn on all plots using dashed lines \cite[respectively]{peredo95:_three_alfven_mach,tsyganenko95:_model_earth}. The black dots at $3.7\,R_E$ show the boundary of the GUMICS$-$4 inner magnetospheric domain. The black circle in the origo of all plots shows the size of the Earth.}
\label{fig:mshorbit}
\end{figure}

Cluster SC3 spent only a little time in the solar wind from December 2002 to May 2003. However, the spacecraft enters the magnetosheath in each orbit (Figure~\ref{fig:mshorbit}). We select intervals when the value of the magnetic field is around 25\,nT. The field should be fluctuating because of the turbulent deflected flow of the shocked solar wind the temperature should be greater than that in the solar wind. The velocity should decrease to values ranging from 100-300\,km/s. The density of the plasma should increase and reach values of 10-20\,cm\textsuperscript{-3}. The narrow band on the omnidirectional CIS HIA ion spectra from the CSA (https://www.cosmos.esa.int/web/csa/csds-quicklook-plots) widens from the solar wind to the magnetosheath. 15--30\,minutes from each bow shock crossing we considered the Cluster SC3 to have entered into the magnetosheath. At the inner magnetopause boundary of the magnetopause the flow speed and the density drop. The magnetic field strength increases and the magnetic field becomes less turbulent than in the magnetosheath. The wide band in the omnidirectional CIS HIA ion spectra disappears, indicating the plasma has undergone heating. 15-30 minutes before (or after) the appearance of these indicators of the magnetopause crossing our intervals end. All intervals containing large data gaps, non-physical jumps in instrument modes or lasting less than four hours are removed. Hence, 74 intervals considered in our final selection (Table~\ref{tab:msh}). 

\begin{figure}[h]
\centering
\includegraphics[width=0.9\textwidth,angle=0]{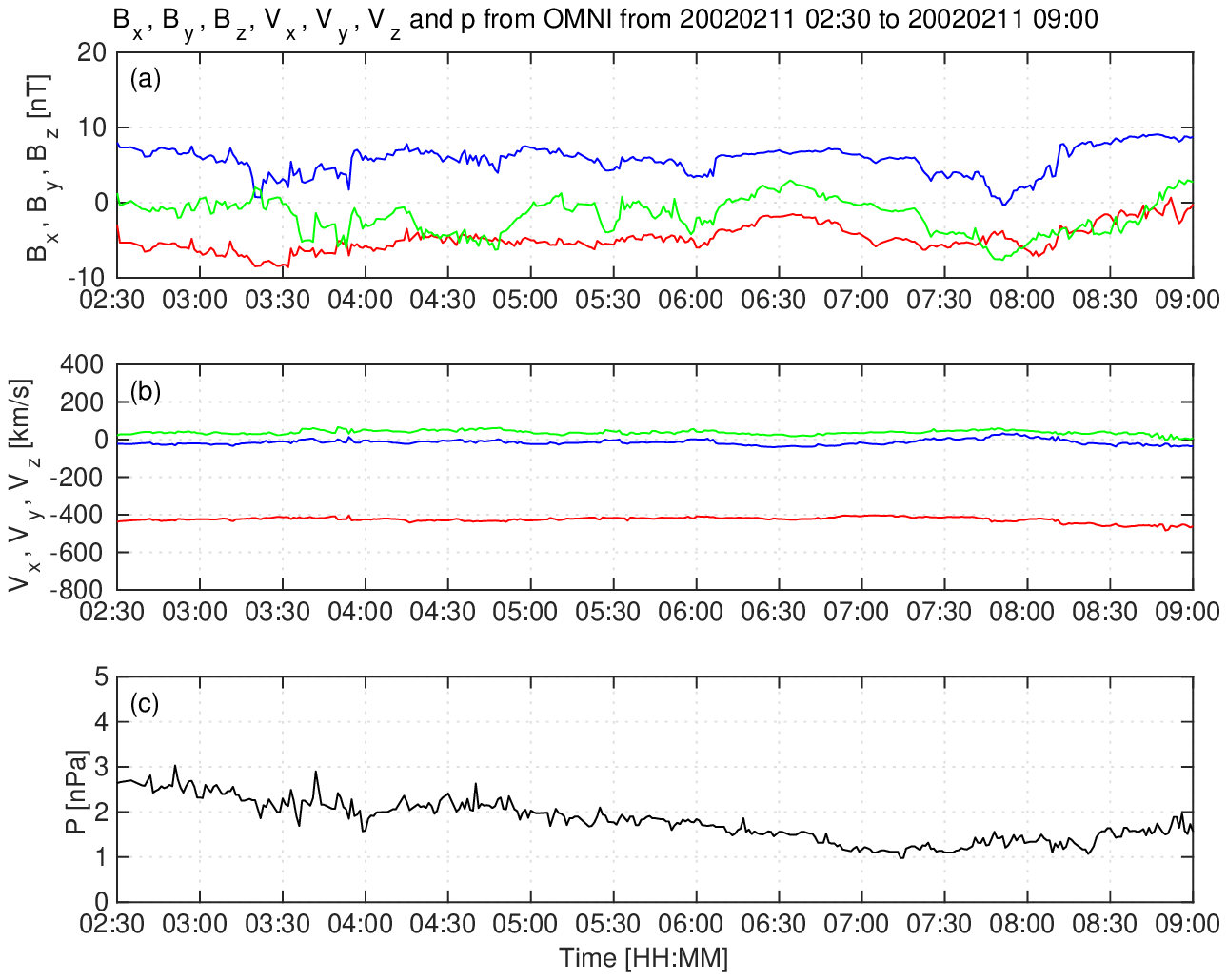} 
\caption{OMNI solar wind data in GSE system from 2:30 to 09:00 (UT) on February 11, 2002. (a) Magnetic field $B_{x}$ (red), $B_{y}$ (green) and $B_{z}$ (blue) components. (b) Solar wind velocity $V_{x}$ (red), $V_{y}$ (green) and $V_{z}$ (blue) components. (c) The $P$ pressure of the solar wind (black).}
\label{fig:mshomni}
\end{figure}

\begin{figure}[h]
\centering
\includegraphics[width=0.9\textwidth,angle=0]{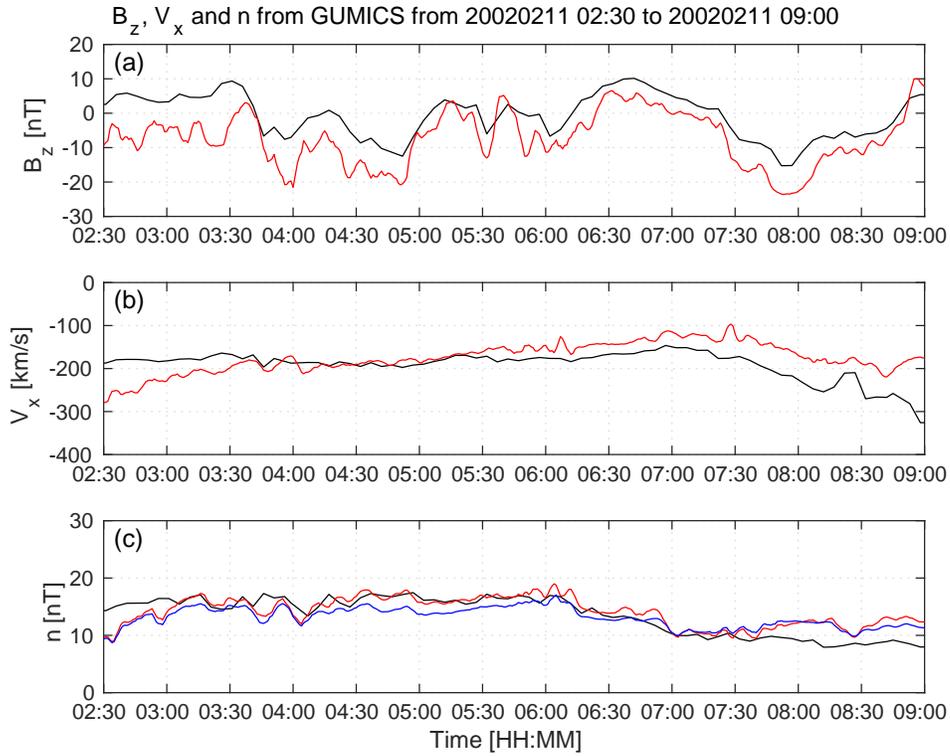} 
\caption{GUMICS-4 simulation results (black) and Cluster SC3 magnetic field Z component, ion plasma moments (red) and electron density calculated from spacecraft potential (blue) from February 11, 2002 from 2:30 to 9:00 (UT) in the magnetosheath in GSE system (a) Magnetic field Z component. (b) Solar wind velocity X component (c) Solar wind density.}
\label{fig:mshplot}
\end{figure}

All intervals have quiet upstream (or input) solar wind conditions (Figure~\ref{fig:mshomni}). Despite our selected quiet magnetic field and plasma parameters, the calculated empirical density indicate that they vary significantly stronger than in the solar wind intervals (Figure~\ref{fig:mshplot}). The deviation between the simulated and the observed data is also larger in this region than in the solar wind region. The scatter plots of the magnetic field, plasma flow speed, and the densities show that these parameters agree well, but with a greater variation than the scatter plots for the same parameters in the solar wind (Figure~\ref{fig:mshscatplot}a,~\ref{fig:mshscatplot}b,~\ref{fig:mshscatplot}c). The correlation of the simulated and the observed data is good for the magnetic field (Figure~\ref{fig:mshcorrplot}a), very good for the ion plasma moments and the calculated density (Figure~\ref{fig:mshcorrplot}c, ~\ref{fig:mshcorrplot}e, ~\ref{fig:mshcorrplot}g). The timeshift of the magnetic field is within five minutes mostly (Figure~\ref{fig:mshcorrplot}b) however the timeshift of the ion plasma moments is scattered (Figure~\ref{fig:mshcorrplot}d,~\ref{fig:mshcorrplot}f). The timeshift of the calculated EFW density seems to be more stable (Figure~\ref{fig:mshcorrplot}h). Generally, the GUMICS$-$4 is less accurate in the magnetosheath than in the solar wind. The modeled magnetic field is better predicted than the modeled plasma parameters are. The calculated empirical EFW density ($n_{EFW}$) fits better than the CIS HIA density ($n_{CIS}$).

\subsection{Magnetosphere}
\label{sec:msph}

\begin{figure}[h]
\centering
\includegraphics[width=0.9\textwidth,angle=0]{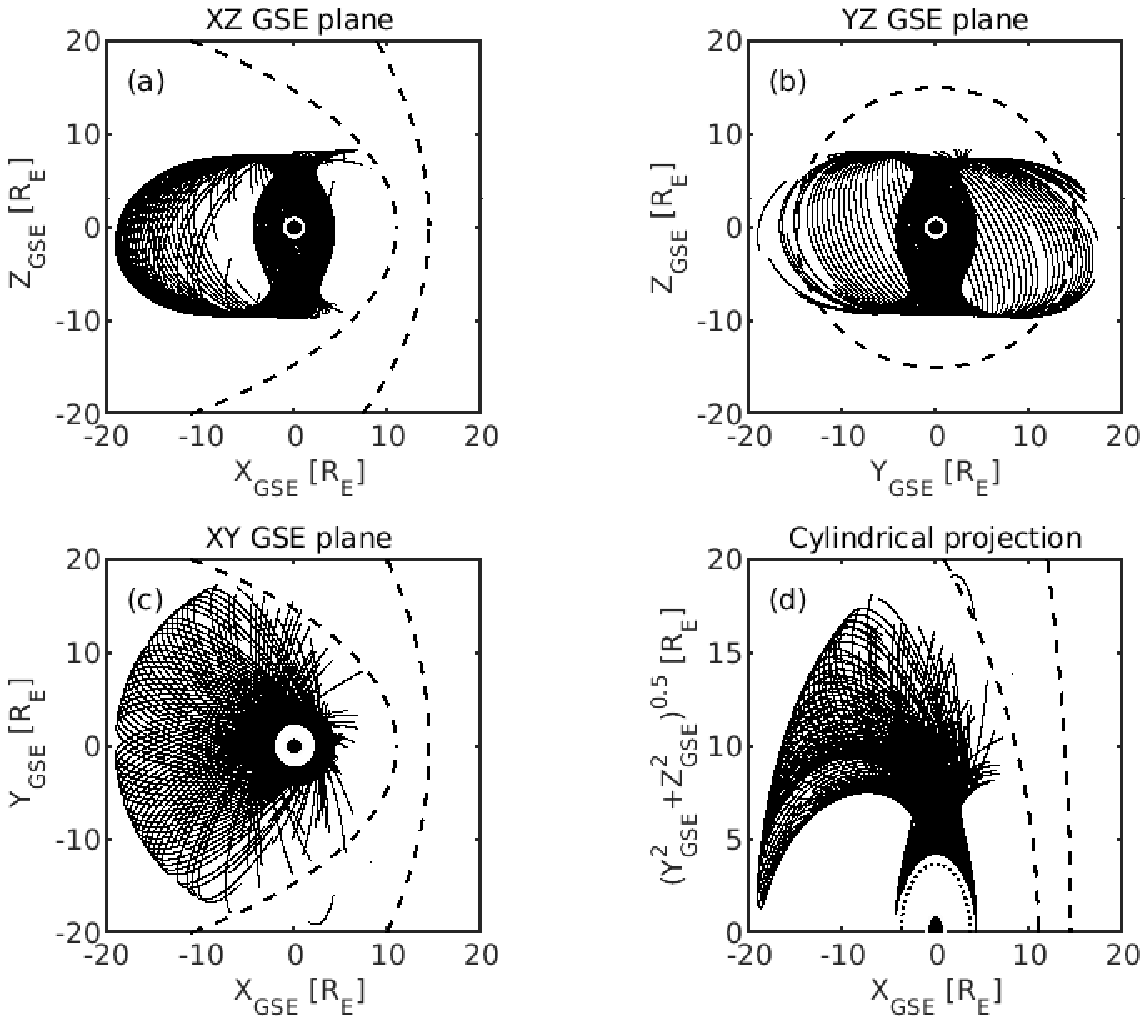} 
\caption{Cluster SC3 orbit in the magnetosphere in GSE system for all intervals (see Table~\ref{tab:msph}). (a) XZ (b) YZ (c) XY (d) Cylindrical projection. Average bow-shock and magnetopause positions are drawn on all plots using dashed lines \cite[respectively]{peredo95:_three_alfven_mach,tsyganenko95:_model_earth}. The black dots at $3.7\,R_E$ show the boundary of the GUMICS$-$4 inner magnetospheric domain. The black circle in the origo of all plots shows the size of the Earth.}
\label{fig:msphorbit}
\end{figure}

We select intervals in the magnetosphere based on the CIS HIA omnidirectional ion flux spectrum. The magnetosphere is defined by the disappearance of hot magnetosheath ion population. The plasma density decreases toward zero, the magnetic field strength becomes great. We eliminated 15--30\,min after/before the magnetopause transition to identify magnetosphere intervals. This way we found 132 intervals of which we found 132 (Table~\ref{tab:msph}) using Cluster SC3 measurements. Cluster SC3 spends considerable time in the magnetosphere (Figure~\ref{fig:msphorbit}). 

Here we show neither any correlation calculation nor comparison plot. In the magnetosphere, the GUMICS$-$4 does not work well. Neither the magnetic field nor the plasma moments nor the $N_{EFW}$ fit well. The solar wind velocity does not reach zero in the simulation. Instead, the solar wind enters the night side magnetosphere. The solar wind CIS HIA ion plasma density and the calculated density from spacecraft potential increase closer to the Earth (plasmasphere). The GUMICS$-$4 density is low there. We calculated the dipole field in GSE using Tsyganenko Geotool box \cite{tsyganenko95:_model_earth} and subtracted from both the observed and the simulated magnetic field $B_z$ data. The correlation of these corrected magnetic field measurements and simulations is very low too. 

\subsection{Bow shock, magnetopause, neutral sheet locations}
\label{sec:bs}

We selected 77 intervals when Cluster SC3 crossed the terrestrial bow shock once or multiple times (Table~\ref{tab:bs}). When the spacecraft crosses the bow shock inbound the magnitude of the magnetic field and the solar wind density increases by a factor to 4--5 (from 5\,nT or 5\,$cm^{-3}$, respectively), the solar wind speed drops from 400--600\,km/s to 100--300\,km/s; furthermore the narrow band in the omnidirectional Cluster CIS HIA ion spectra widens. Both the Cluster measurements and the GUMICS$-$4 simulations have 5--min resolution and are interpolated to 1--min resolution. All bow shock transitions of the virtual spacecraft are slower and smoother. Additionally, GUMICS-4 does not predict multiple bow shock transitions. The code reacts slowly to such sudden changes. The magnetic signatures fit better than the calculated plasma moments. The jump of the ion plasma parameters and the derived Cluster EFW density of the simulations are shifted to the measurements. Generally, the density and the velocity of the simulations seem to be less accurate than the magnetic field in the simulations.

54 intervals are selected around magnetopause crossings (Table~\ref{tab:mp}). When the spacecraft crosses the magnetopause inbound the magnitude of the magnetic field increases, the solar wind speed drops from 100--300\,km/s to zero, the plasma density becomes zero; furthermore the wide band on the omnidirectional Cluster CIS HIA ion spectra disappears. The location of the magnetopause is well determined by the Cluster SC3 measurements. However, it is very difficult to identify the magnetopause crossings in the simulation data. The magnetopause crossings usually (92\,\%) cannot be seen in the simulations. The magnetopause crossings are not visible in $V_{x}$ and n. This observation is independent of the IMF orientation. Or when the magnetopause crossings are identified in both simulations and spacecraft measurements the events are shifted in time and location. The accuracy of the model is lower for the dayside magnetopause locations than the bow shock locations. 

\begin{figure}[h]
\centering
\includegraphics[width=0.9\textwidth,angle=0]{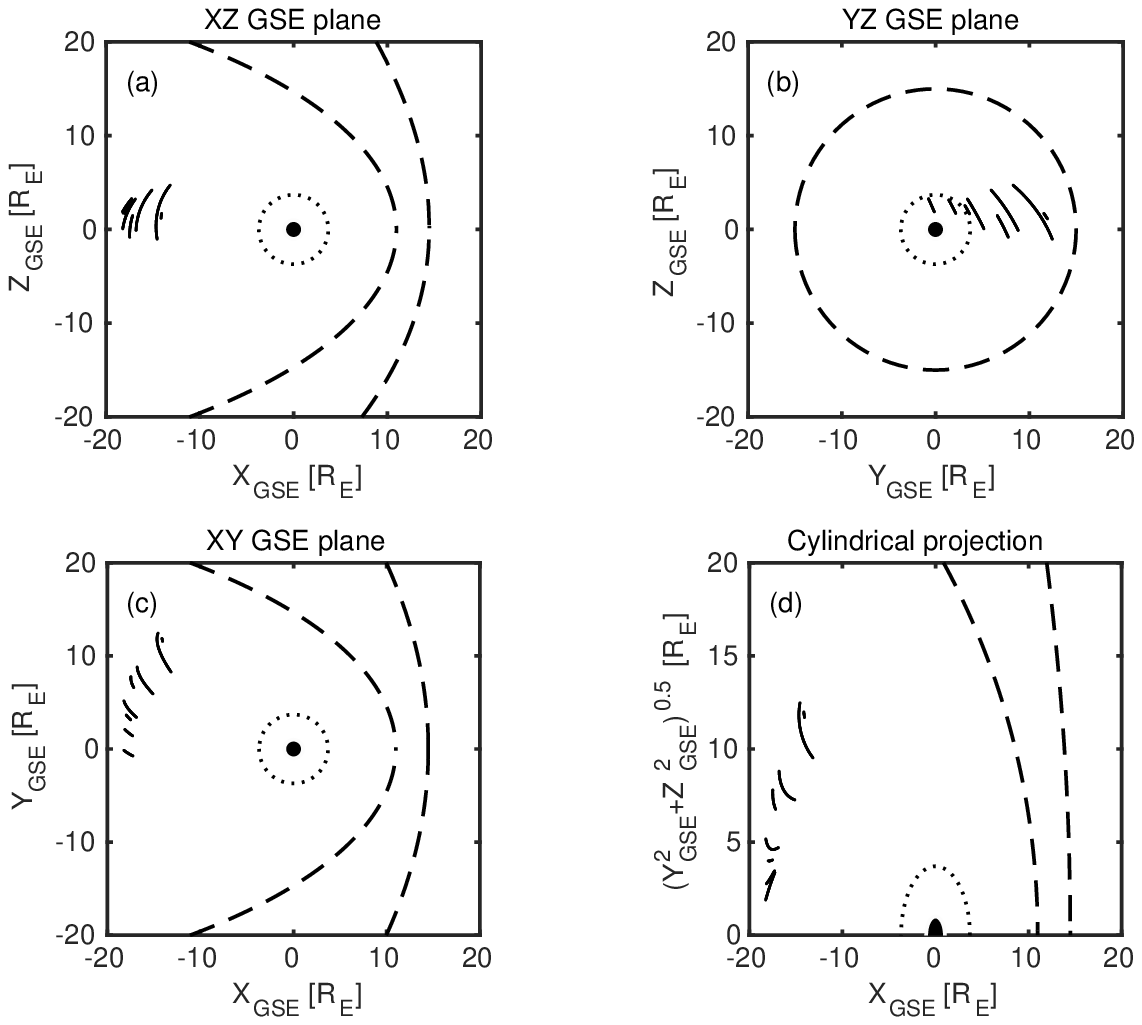} 
\caption{Cluster SC3 orbit in the tail in GSE system for all intervals (see Table~\ref{tab:ns}). (a) XZ (b) YZ (c) XY (d) Cylindrical projection. Average bow-shock and magnetopause positions are drawn on all plots using dashed lines \cite[respectively]{peredo95:_three_alfven_mach,tsyganenko95:_model_earth}. The black dots at $3.7\,R_E$ show the boundary of the GUMICS$-$4 inner magnetospheric domain. The black circle in the origo of all plots shows the size of the Earth.}
\label{fig:nsorbit}
\end{figure}

\begin{figure}[h]
\centering
\includegraphics[width=0.9\textwidth,angle=0]{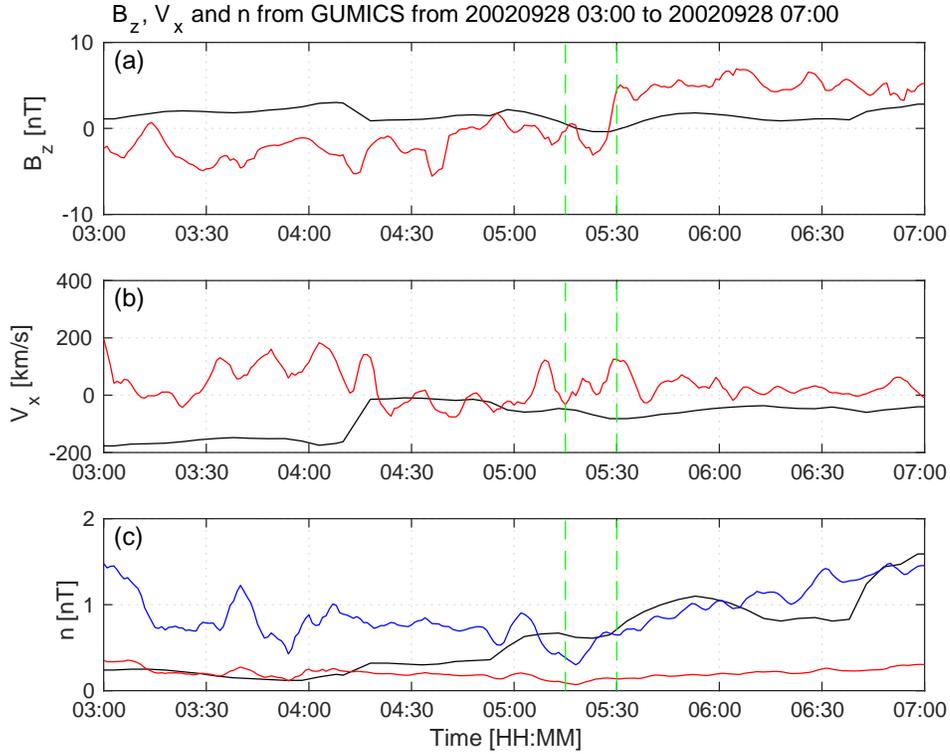} 
\caption{GUMICS-4 simulation results (black) and Cluster SC3 magnetic field Z component, ion plasma moments (red) and electron density calculated from spacecraft potential (blue) from September 28, 2002 from 3:00 to 7:00 (UT) in the tail in GSE system. (a) Magnetic field Z component. (b) Solar wind velocity X component (c) Solar wind density. From 05:15 to 05:30 between the green dashed vertical lines both the Cluster SC3 and the virtual spaceprobe of the GUMICS$-$4 simulation cross the neutral sheet multiple times.}
\label{fig:nsplot}
\end{figure}

Nine intervals are chosen around Cluster SC3 neutral sheet crossings (Figure~\ref{fig:nsorbit}; Table~\ref{tab:ns}). The neutral sheet locations are determined using the results of the Boundary Layer Identification Code (BLIC) Project (E.~I.~Tanskanen, private communication, 2015). The BLIC code determines the neutral sheet crossing Cluster FGM magnetic field measurements using \citeA{wang94:_signat}'s method. When the solar wind speed is very low around the currents sheet in the simulation space; furthermore the CIS HIA density and the EFW calculated density are very low near the current sheet too; finally the GSE Z component of the magnetic field changing is a sign of the code--indicated neutral sheet crossing (Figure~\ref{fig:nsplot}; red and blue curves). The neutral sheet crossings are visible in the GUMICS simulations (Figure~\ref{fig:nsplot}; black curves). For five events (from nine Cluster SC3 crossings) the GUMICS$-$4 also provides similar smoothed parameters and change of sign of the $B_{z}$ component. This is an outstanding result because the tail in the GUMICS$-$4 simulations is significantly smaller than observed in reality \cite{gordeev13:_verif_gumic_mhd,facsko16:_one_earth}; furthermore the solar wind enters the tail in MHD simulations generally \cite{kallio15:_proper}.

\section{Discussion}
\label{sec:discussion}

\begin{figure}[h]
\centering
\includegraphics[width=0.9\textwidth,angle=0]{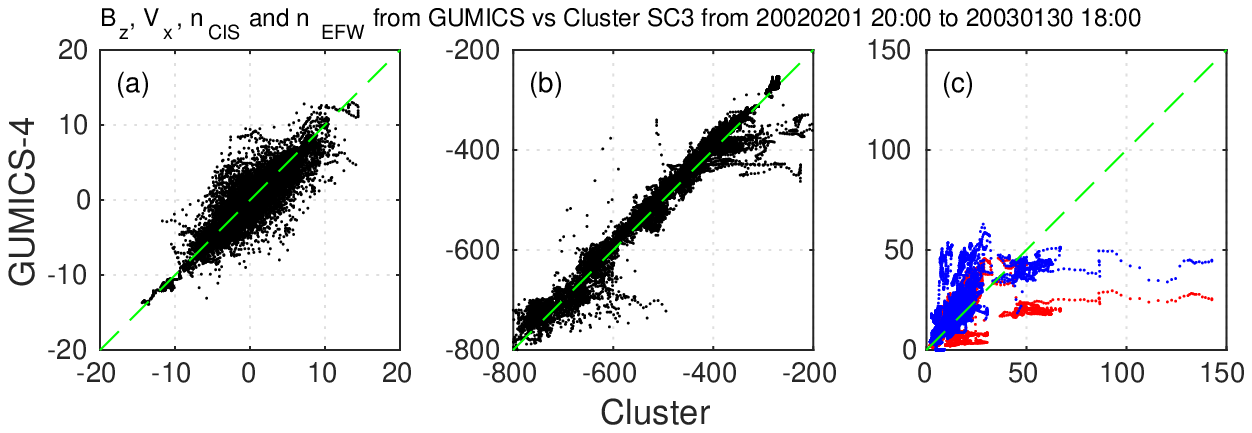}
\caption{Scattered plots of the Cluster SC3 and GUMICS$-$4 simulations for all intervals in the solar wind. The dashed line is the y=x line. (a) Magnetic field Z component in GSE system. (b) Solar wind velocity X component in GSE system. (c) Solar wind density measured by the CIS HIA instrument (red) and calculated from the spacecraft potential (blue).}
\label{fig:swscatplot}
\end{figure}

The agreement of $B_{z}$, $V_{x}$ and $n_{EFW}$ in the solar wind with the similar GUMICS simulation predictions is very good (Figure~\ref{fig:swscatplot}a,~\ref{fig:swscatplot}b,~\ref{fig:swscatplot}c,~blue). The agreement of $n_{CIS}$ is worse (Figure~\ref{fig:swscatplot}c,~red). It was expected because the $n_{EFW}$ depends on the spacecraft potential provided by the EFW instrument. However, the CIS instrument has many modes for measuring the plasma parameters and it needs periodic calibration too. The correlation of the solar wind $V_{x}$, $n_{CIS}$ and $n_{EFW}$ with the similar GUMICS simulation parameters is greater than 0.9 (Figure~\ref{fig:swcorrplot}c,~\ref{fig:swcorrplot}e,~\ref{fig:swcorrplot}g). The correlation of the $B_{z}$ is also greater than 0.8 (Figure~\ref{fig:swcorrplot}a). The upstream boundary of the GUMICS$-$4 code lies at $32\,R_E$ \cite{janhunen12:_gumic_mhd}, the nose of the terrestrial bow shock is at about $20\,R_E$. If the solar wind speed is 400\,km/s, then this spatial distance means less than a 5\,minutes delay, so it should not be visible in the time delays from the cross--correlations. 80\% of the intervals support this theory but 20\,\% do not. In these cases, the one-minute resolution $B_z$, $n_{CIS}$ or the $n_{EFW}$ parameters have a sudden jump or variation that the simulation cannot follow, or the resolution of the simulation output (5\,minutes) is too small to see these variations. Therefore, the correlation calculation is not accurate in these cases. Previously the OMNI data was compared to the Cluster data and the Cluster measurements were compared to the GUMICS$-$4 \cite{facsko16:_one_earth}. The comparison suggests that the GUMICS$-$4 results should be similar to the OMNI data. Furthermore, we calculate correlation functions in the solar wind, where there is no significant perturbation of the input parameters in the simulation box. Therefore, we get the expected result after comparing the two different correlation calculations.

\begin{figure}[h]
\centering
\includegraphics[width=0.9\textwidth,angle=0]{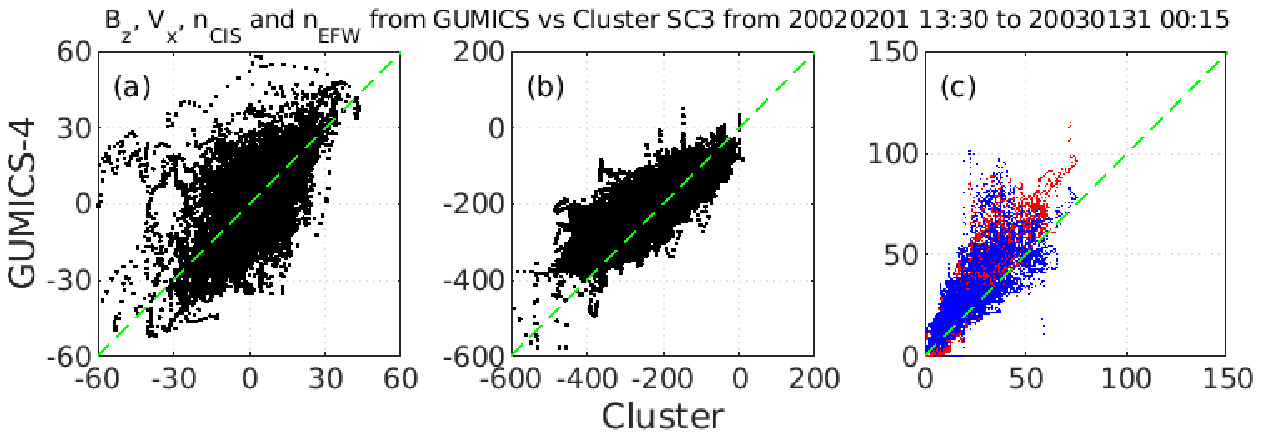}
\caption{Scattered plots of the Cluster SC3 and GUMICS$-$4 simulations for all intervals in the magnetosheath in GSE system. The dashed line is the y=x line. (a) Magnetic field Z component. (b) Solar wind velocity X component. (c) Solar wind density measured by the CIS HIA instrument (red) and calculated from the spacecraft potential (blue).}
\label{fig:mshscatplot}
\end{figure}

\begin{figure}[h]
\centering
\includegraphics[width=0.9\textwidth,angle=0]{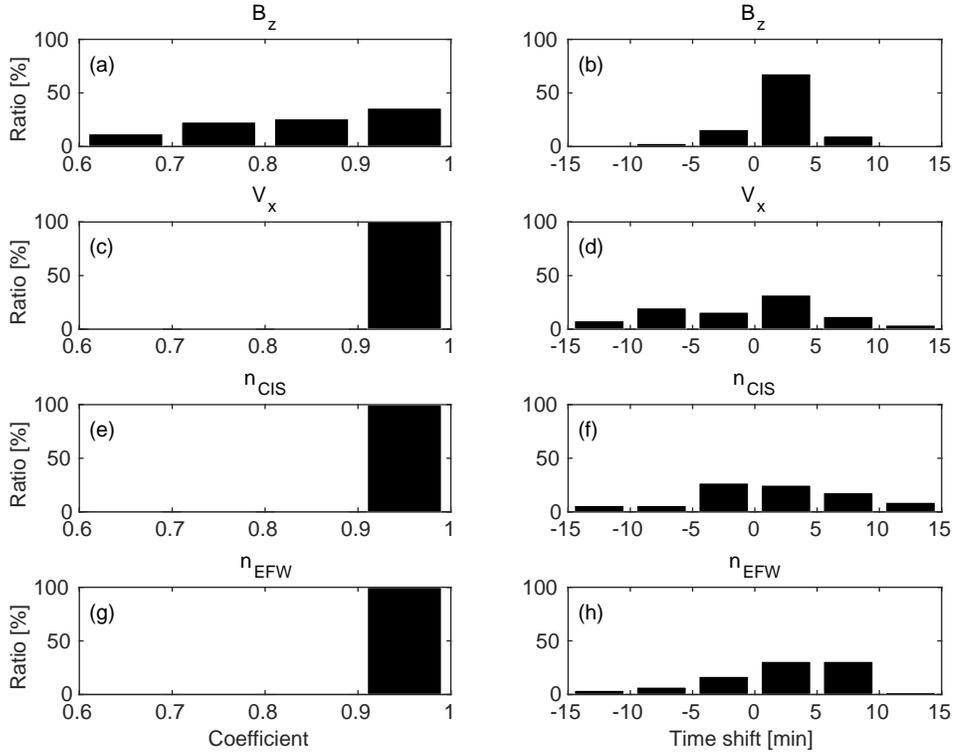}
\caption{The distributions of the cross-correlation coefficients (a,~c,~e,~g) of the magnetic field Z component ($B_z$) in GSE system, solar wind velocity X component ($V_X$) in GSE system, the solar wind density measured by the CIS HIA ($n_{CIS}$) instrument and calculated from the spacecraft potential ($n_{EFW}$), respectively, for all intervals in the magnetosheath. The distributions of the time shifts (b,~d,~f,~h) of the $B_z$, the $V_X$, the $n_{CIS}$ and the $n_{EFW}$), respectively, for all intervals in the magnetosheath.}
\label{fig:mshcorrplot}
\end{figure}

In the magnetosheath we get worse agreement with the GUMICS simulation data (Figure~\ref{fig:mshscatplot}a,~\ref{fig:mshscatplot}b,~\ref{fig:mshscatplot}c). While the parameters are correlated, the scatter is greater. The general reason for this larger uncertainty seems to be that the magnetosheath is turbulent. This phenomenon explains the higher variations of the $B_{z}$ magnetic field on Figure~\ref{fig:mshscatplot}a. The solar wind $V_{x}$, $n_{CIS}$ and $n_{EFW}$ agree better than the magnetic field component (Figure~\ref{fig:mshscatplot}b,~\ref{fig:mshscatplot}c). Here there is no deviation between the densities derived in different ways ($n_{CIS}$ and $n_{EFW}$) on Figure~\ref{fig:mshscatplot}c. Figure~\ref{fig:mshcorrplot} seems to contradict these statements above. The larger uncertainty of the $B_{z}$ is visible in Figure~\ref{fig:mshcorrplot}a. However, that correlation is still good in Figure~\ref{fig:mshcorrplot}b. The other parameters have larger ($>0.9$) correlation in Figure~\ref{fig:mshcorrplot}c,~\ref{fig:mshcorrplot}e,~\ref{fig:mshcorrplot}g. However, the time shifts in Figure~\ref{fig:mshcorrplot}d,~\ref{fig:mshcorrplot}f,~\ref{fig:mshcorrplot}h seem to be worse. Here the time shifts are worse because the shape of the time series in the magnetosheath looks very smooth and similar hence there are not enough points to get a sharp and large maximum correlation as the function of timeshift. The difference between the minimum and the maximum of the correlation is small compared with the uncertainty of the calculation. The maximum, the timeshift could be anywhere and the shape of the correlation vs.~timeshift function is often neither symmetric nor does it have only one local maximum. Hence, the correlation calculation provides larger time shifts for the ion plasma parameters and the $n_{EFW}$. 

In the magnetosphere, the GUMICS$-$4 does not work well. GUMICS$-$4 uses a tilted dipole to describe the terrestrial magnetic field \cite{janhunen12:_gumic_mhd}. After removing the magnetic dipole from the magnetic field measurements of the Cluster SC3 and the simulation we get very low correlations and unacceptable time shifts (not shown). The tilted dipole is an insufficient description of the inner magnetospheric magnetic field. The plasma moments and the $n_{EFW}$ do not fit either. The single fluid, ideal MHD does not describe the inner magnetosphere well therefore $V_{x}$ and $n$ in the simulations do not agree with $V_{x}$, $n_{CIS}$ and the $n_{EFW}$ measured by the Cluster SC3. Within the $3.7\,R_{E}$ domain ring current physics must be added, as it has been in other global MHD codes \cite<for example>[]{toth12:_adapt}. This can explain the limited accuracy of the cross--polar cap potential (CPCP) and geomagnetic indices of the GUMICS simulations \cite{juusola14:_statis_gumic_mhd}. The CPCP GUMICS agrees well with spacecraft measurements therefore this quantity could be used for simulation studies \cite{lakka18:_cross_polar_cap_satur_gumic}. \citeA{haiducek17:_swmf_global_magnet_simul_januar} also compared geomagnetic indices and the CPCP. The Space Weather Modelling Framework (SWMF) was tested. When the inner magnetosphere model was switched off in the simulation only the comparison of the simulated and observed CPCP was good. Therefore, the reason for the discrepancy of the geomagnetic indices in the GUMICS simulations must be the missing inner magnetosphere model.

\begin{figure}[h]
\centering
\includegraphics[width=0.6\textwidth,angle=0]{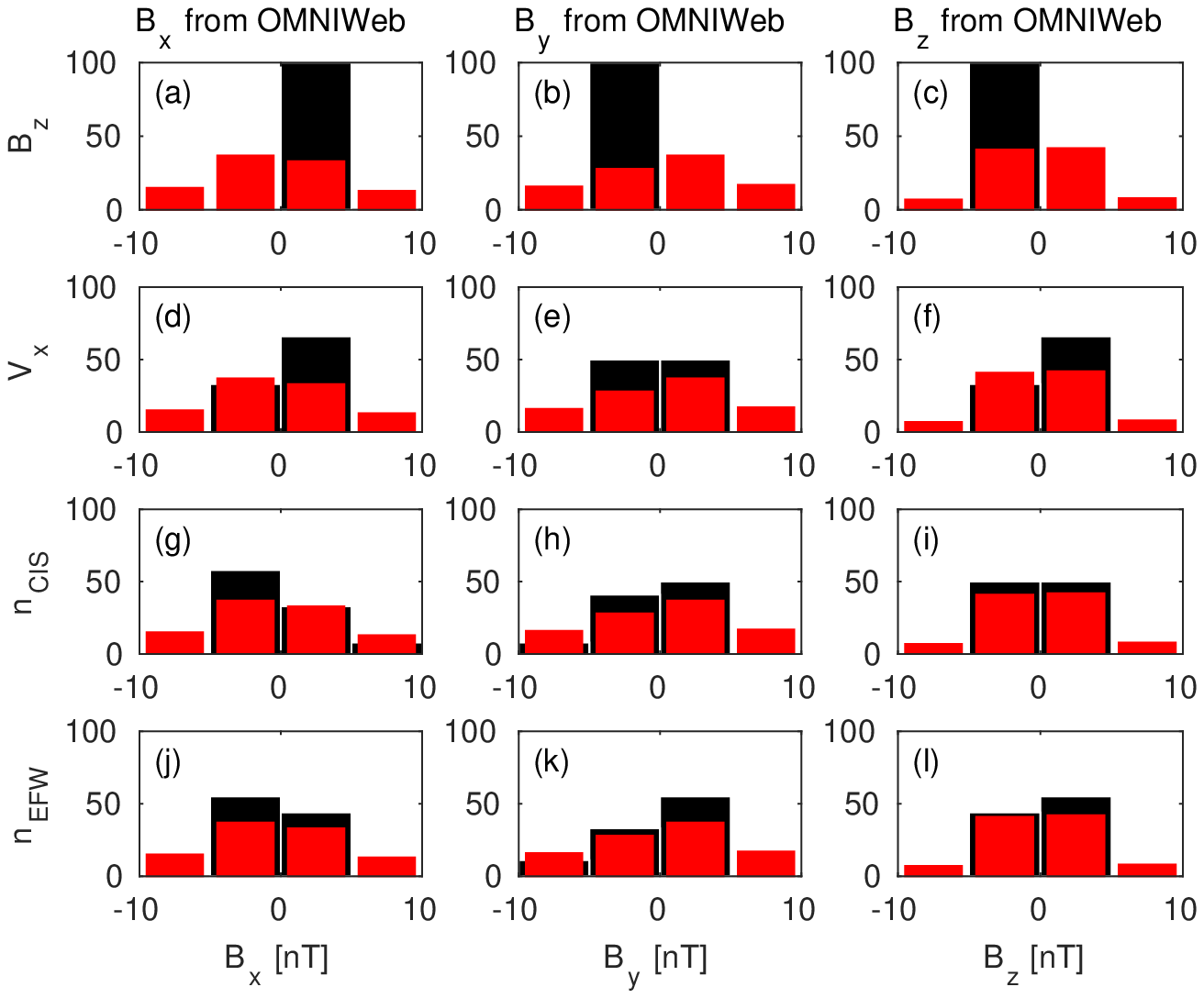} 
\caption{The black distributions of the $B_{x}$, the $B_{y}$ and the $B_{z}$ OMNI solar wind magnetic field components when the agreement of the Cluster SC3 measurements and the GUMICS$-$4 simulations are poor in the solar wind (see Table~\ref{tab:omnisw}). The $B_{z}$, the $V_{x}$, the $n_{CIS}$ and the $n_{EFW}$ are the magnetic field GSE Z component, the plasma ion velocity X GSE component, the solar wind density measured by the CIS HIA instrument and the calculated from the EFW spacecraft potential, respectively. (a,~b,~c) Distribution of OMNI $B_{x}$, $B_{y}$, $B_{z}$ when the agreement of $B_{z}$ is poor. (d,~e,~f) Distribution of OMNI $B_{x}$, $B_{y}$, $B_{z}$ when the agreement of $V_{x}$ is poor. (g,~h,~i) Distribution of OMNI $B_{x}$, $B_{y}$, $B_{z}$ when the agreement of $n_{CIS}$ is poor. (j,~k,~l) Distribution of OMNI $B_{x}$, $B_{y}$, $B_{z}$ when the agreement of $n_{EWF}$ is poor. The values are in percentage unitss in the distributions. The red distributions of (a,~d,~g,~j), (b,~e,~h,~k) and (c,~f,~i,~l) are the distribution of the $B_{x}$, the $B_{y}$, and the $B_{z}$ components of the OMNI solar wind magnetic field during the 1-year run from January 29, 2002, to February 2, 2003, in GSE reference frame, respectively.} 
\label{fig:swomnibxyz}
\end{figure}

\begin{figure}[h]
\centering
\includegraphics[width=0.6\textwidth,angle=0]{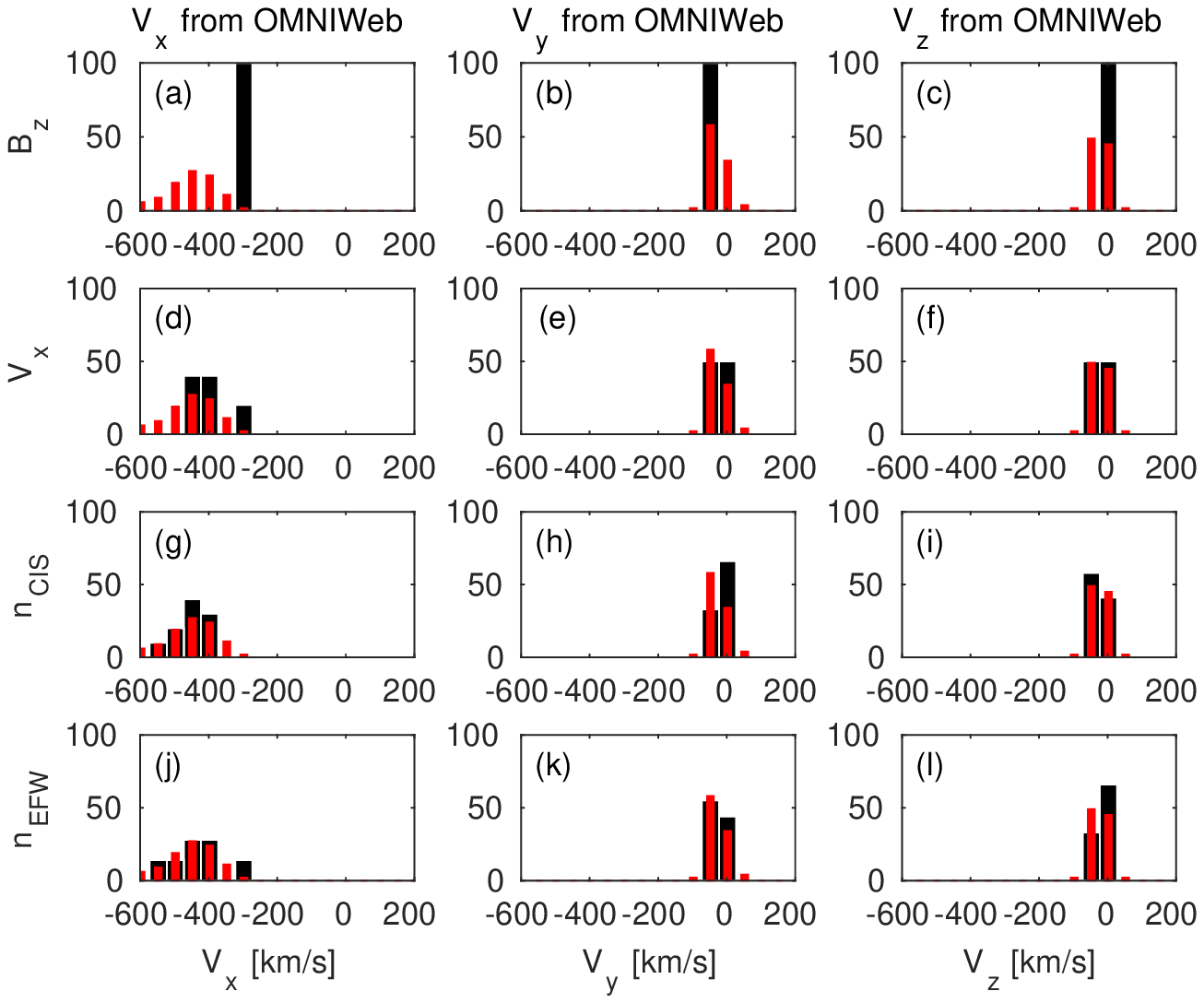}
\caption{The black distributions of the $V_{x}$, the $V_{y}$ and the $V_{z}$ OMNI solar wind velocity components when the agreement of the Cluster SC3 measurements and the GUMICS$-$4 simulations are poor in the solar wind (see Table~\ref{tab:omnisw}). The $B_{z}$, the $V_{x}$, the $n_{CIS}$ and the $n_{EFW}$ are the magnetic field GSE Z component, the plasma ion velocity X GSE component, the solar wind density measured by the CIS HIA instrument and the calculated from the EFW spacecraft potential, respectively. (a,~b,~c) Distribution of OMNI $V_{x}$, $V_{y}$, $V_{z}$ when the agreement of $B_{z}$ is poor. (d,~e,~f) Distribution of OMNI $V_{x}$, $V_{y}$, $V_{z}$ when the agreement of $V_{x}$ is poor. (g,~h,~i) Distribution of OMNI $V_{x}$, $V_{y}$, $V_{z}$ when the agreement of $n_{CIS}$ is poor. (j,~k,~l) Distribution of OMNI $V_{x}$, $V_{y}$, $V_{z}$ when the agreement of $n_{EWF}$ is poor. The values are in percentage units in the distributions. The red distributions of (a,~d,~g,~j), (b,~e,~h,~k) and (c,~f,~i,~l) are the distributions of the $V_{x}$, the $V_{y}$ and the $V_{z}$ components of the OMNI solar wind velocity during the 1-year run from January 29, 2002 to February 2, 2003 in GSE reference frame, respectively.}
\label{fig:swomnivxyz}
\end{figure}

\begin{figure}[h]
\centering
\includegraphics[width=0.7\textwidth,angle=0]{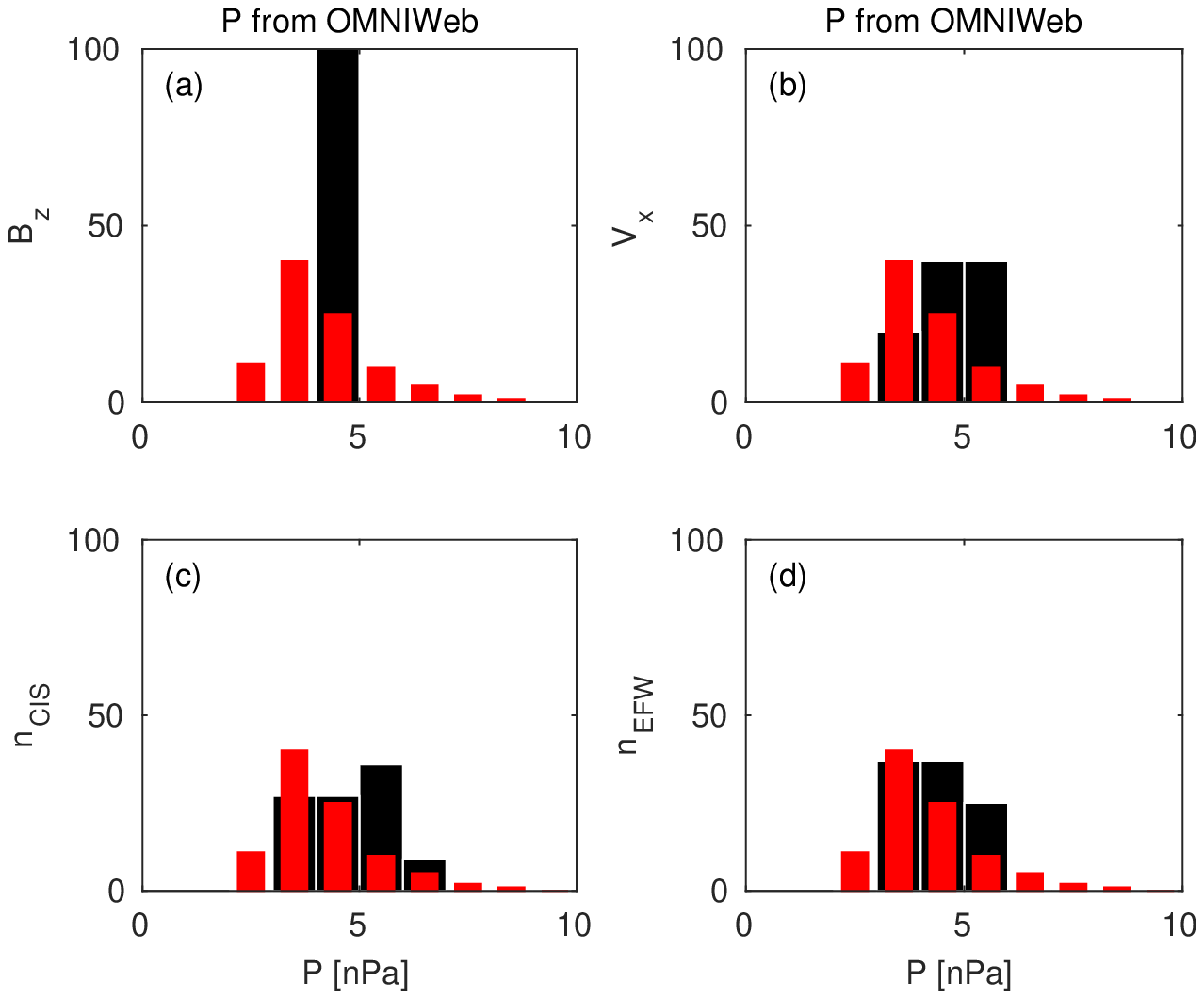} 
\caption{The black distributions of the $P$ solar wind dynamic pressure calculated from OMNI parameters when the agreement of the Cluster SC3 measurements and the GUMICS$-$4 simulations are poor in the solar wind (see Table~\ref{tab:omnisw}). The $B_{z}$, $V_{x}$, $n_{CIS}$ and $n_{EFW}$ are the magnetic field GSE Z component, the velocity X GSE component, the solar wind density measured by the CIS HIA instrument and calculated from the EFW spacecraft potential, respectively. (a,~b,~c,~d) The distribution of the P calculated from OMNI data when the agreement of the $B_{z}$, the $V_{x}$, the $n_{CIS}$ or the $n_{EFW}$ are poor. The values are in percentage units in the distributions. The red distributions of (a,~b,~c,~d) are the distributions of the P solar wind dynamic pressure calculated from the OMNI solar wind parameters during the 1-year run from January 29, 2002, to February 2, 2003, in GSE reference frame.}
\label{fig:swomnip}
\end{figure}

\begin{figure}[h]
\centering
\includegraphics[width=0.7\textwidth,angle=0]{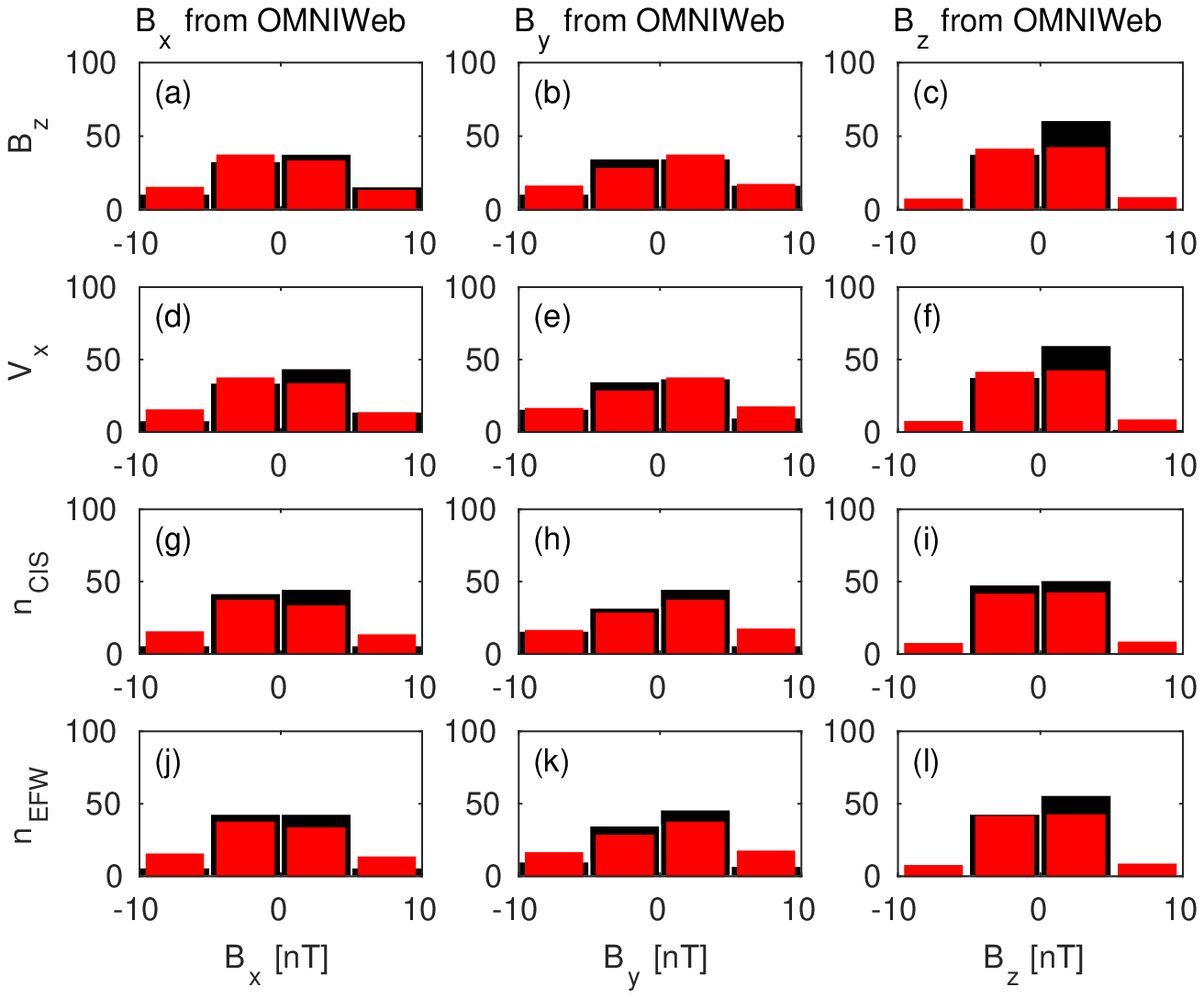}
\caption{The black distributions of the $B_{x}$, the $B_{y}$ and the $B_{z}$ OMNI solar wind magnetic field components when the agreement of the Cluster SC3 measurements and the GUMICS$-$4 simulations are poor in the magnetosheath (see Table~\ref{tab:omnimsh}). The $B_{z}$, the $V_{x}$, the $n_{CIS}$ and the $n_{EFW}$ are the magnetic field GSE Z component, the plasma ion velocity X GSE component, the solar wind density measured by the CIS HIA instrument and the calculated from the EFW spacecraft potential, respectively. (a,~b,~c) Distribution of OMNI $B_{x}$, $B_{y}$, $B_{z}$ when the agreement of $B_{z}$ is poor. (d,~e,~f) Distribution of OMNI $B_{x}$, $B_{y}$, $B_{z}$ when the agreement of $V_{x}$ is poor. (g,~h,~i) Distribution of OMNI $B_{x}$, $B_{y}$, $B_{z}$ when the agreement of $n_{CIS}$ is poor. (j,~k,~l) Distribution of OMNI $B_{x}$, $B_{y}$, $B_{z}$ when the agreement of $n_{EWF}$ is poor. The values are in percentage units in the distributions. The red distributions of (a,~d,~g,~j), (b,~e,~h,~k) and (c,~f,~i,~l) are the distribution of the $B_{x}$, the $B_{y}$, and the $B_{z}$ components of the OMNI solar wind magnetic field during the 1-year run from January 29, 2002, to February 2, 2003, in GSE reference frame, respectively.}
\label{fig:mshomnibxyz}
\end{figure}

\begin{figure}[h]
\centering
\includegraphics[width=0.7\textwidth,angle=0]{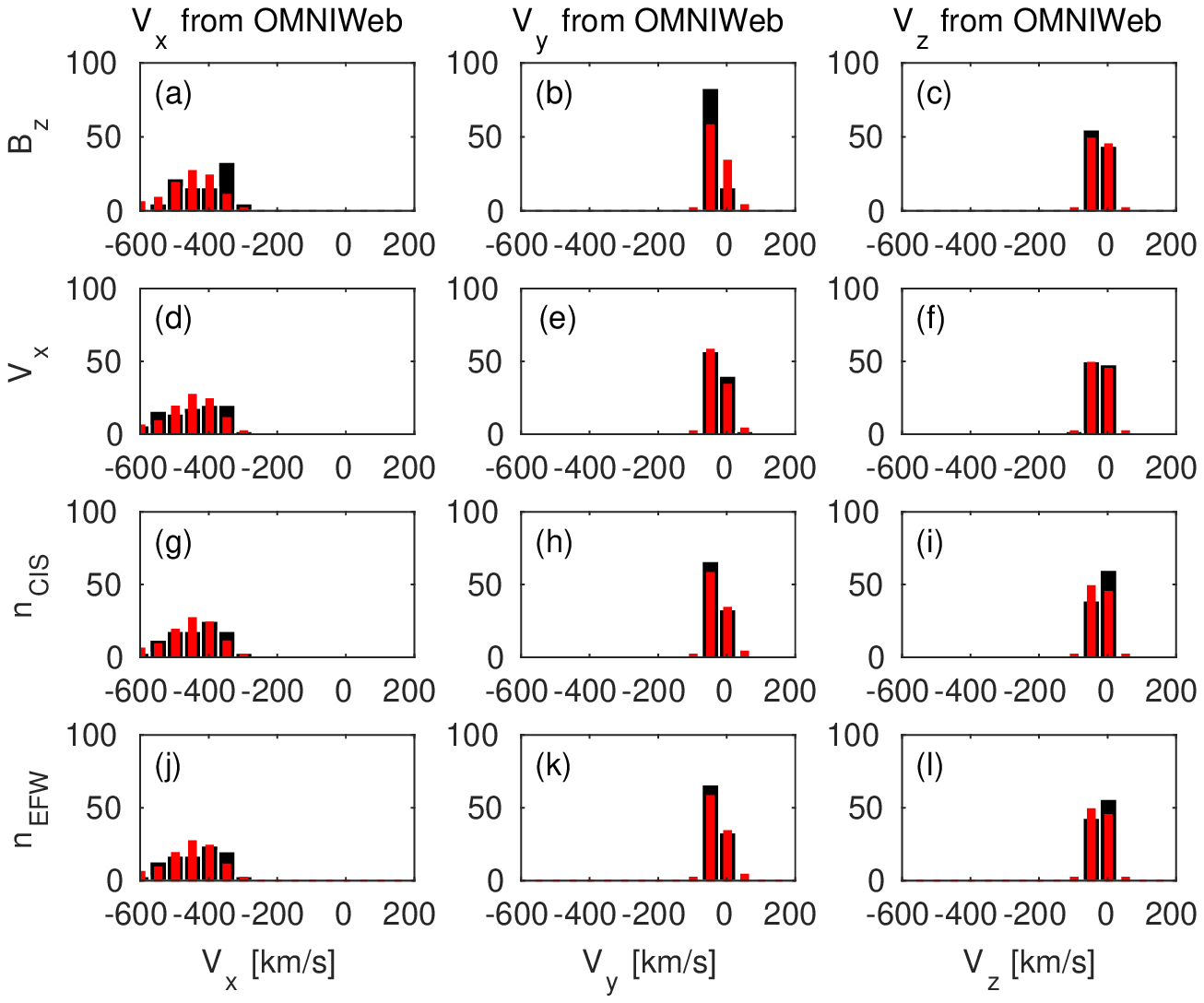}
\caption{The black distributions of the $V_{x}$, the $V_{y}$ and the $V_{z}$ OMNI solar wind velocity components when the agreement of the Cluster SC3 measurements and the GUMICS$-$4 simulations are poor in the magnetosheath (see Table~\ref{tab:omnimsh}). The $B_{z}$, the $V_{x}$, the $n_{CIS}$ and the $n_{EFW}$ are the magnetic field GSE Z component, the plasma ion velocity X GSE component, the solar wind density measured by the CIS HIA instrument and the calculated from the EFW spacecraft potential, respectively. (a,~b,~c) Distribution of OMNI $V_{x}$, $V_{y}$, $V_{z}$ when the agreement of $B_{z}$ is poor. (d,~e,~f) Distribution of OMNI $V_{x}$, $V_{y}$, $V_{z}$ when the agreement of $V_{x}$ is poor. (g,~h,~i) Distribution of OMNI $V_{x}$, $V_{y}$, $V_{z}$ when the agreement of $n_{CIS}$ is poor. (j,~k,~l) Distribution of OMNI $V_{x}$, $V_{y}$, $V_{z}$ when the agreement of $n_{EWF}$ is poor. The values are in percentage units in the distributions. The red distributions of (a,~d,~g,~j), (b,~e,~h,~k) and (c,~f,~i,~l) are the distributions of the $V_{x}$, the $V_{y}$ and the $V_{z}$ components of the OMNI solar wind velocity during the 1-year run from January 29, 2002 to February 2, 2003 in GSE reference frame, respectively.}
\label{fig:mshomnivxyz}
\end{figure}

\begin{figure}[h]
\centering
\includegraphics[width=0.9\textwidth,angle=0]{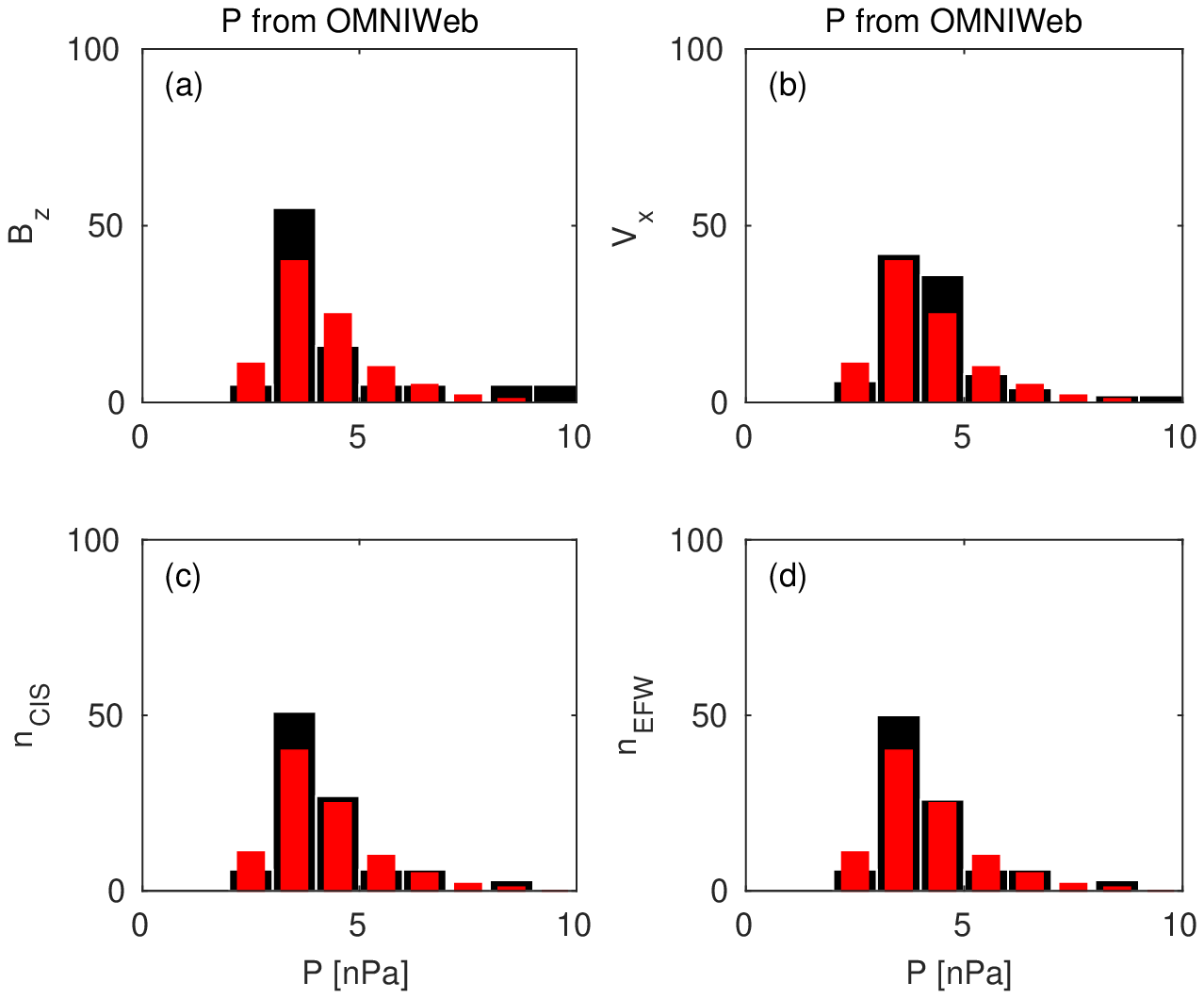}
\caption{The black distributions of the $P$ solar wind dynamic pressure calculated from OMNI parameters when the agreement of the Cluster SC3 measurements and the GUMICS$-$4 simulations are poor in the magnetosheath (see Table~\ref{tab:omnimsh}). The $B_{z}$, $V_{x}$, $n_{CIS}$ and $n_{EFW}$ are the magnetic field GSE Z component, the velocity X GSE component, the solar wind density measured by the CIS HIA instrument and calculated from the EFW spacecraft potential, respectively. (a,~b,~c,~d) The distribution of the P calculated from OMNI data when the agreement of the $B_{z}$, the $V_{x}$, the $n_{CIS}$ or the $n_{EFW}$ are poor. The values are in percentage units in the distributions. The red distributions of (a,~b,~c,~d) are the distributions of the P solar wind dynamic pressure calculated from the OMNI solar wind parameters during the 1-year run from January 29, 2002 to February 2, 2003 in GSE reference frame.}
\label{fig:mshomnip}
\end{figure}

The reason why simulation results and measurements disagree could be the code or bad input parameters. During the 1-year run the distributions of the OMNI solar wind magnetic field $B_{x}$, $B_{y}$, $B_{z}$ components; solar wind velocity $V_{x}$, $V_{y}$ $V_{z}$ components and the solar wind $P$ dynamic pressure are calculated from January 29, 2002 to February 2, 2003 in GSE reference frame. The distributions of the OMNI solar wind magnetic field $B_{x}$, $B_{y}$, $B_{z}$ components were overplotted by red in Figure~\ref{fig:swomnibxyz}a,~\ref{fig:swomnibxyz}d,~\ref{fig:swomnibxyz}g,~\ref{fig:swomnibxyz}j and Figure~\ref{fig:mshomnibxyz}a,~\ref{fig:mshomnibxyz}d,~\ref{fig:mshomnibxyz}g,~\ref{fig:mshomnibxyz}j; Figure~\ref{fig:swomnibxyz}b,~\ref{fig:swomnibxyz}e,~\ref{fig:swomnibxyz}h,~\ref{fig:swomnibxyz}k and Figure~\ref{fig:mshomnibxyz}b,~\ref{fig:mshomnibxyz}e,~\ref{fig:mshomnibxyz}h,~\ref{fig:mshomnibxyz}k; furthermore Figure~\ref{fig:swomnibxyz}c,~\ref{fig:swomnibxyz}f,~\ref{fig:swomnibxyz}i,~\ref{fig:swomnibxyz}l and Figure~\ref{fig:mshomnibxyz}c,~\ref{fig:mshomnibxyz}f,~\ref{fig:mshomnibxyz}i,~\ref{fig:mshomnibxyz}l. The distributions of the OMNI solar wind velocity $V_{x}$, $V_{y}$, $V_{z}$ components were overplotted by red in Figure~\ref{fig:swomnivxyz}a,~\ref{fig:swomnivxyz}d,~\ref{fig:swomnivxyz}g,~\ref{fig:swomnivxyz}j and Figure~\ref{fig:mshomnivxyz}a,~\ref{fig:mshomnivxyz}d,~\ref{fig:mshomnivxyz}g,~\ref{fig:mshomnivxyz}j; Figure~\ref{fig:swomnivxyz}b,~\ref{fig:swomnivxyz}e,~\ref{fig:swomnivxyz}h,~\ref{fig:swomnivxyz}k and Figure~\ref{fig:mshomnivxyz}b,~\ref{fig:mshomnivxyz}e,~\ref{fig:mshomnivxyz}h,~\ref{fig:mshomnivxyz}k; furthermore Figure~\ref{fig:swomnivxyz}c,~\ref{fig:swomnivxyz}f,~\ref{fig:swomnivxyz}i,~\ref{fig:swomnivxyz}l and Figure~\ref{fig:mshomnivxyz}c,~\ref{fig:mshomnivxyz}f,~\ref{fig:mshomnivxyz}i,~\ref{fig:mshomnivxyz}l. The distributions of the P solar wind pressure calculated from the OMNI solar wind parameters were overplotted by red in Figure~\ref{fig:swomnip}a,~\ref{fig:swomnip}b,~\ref{fig:swomnip}c,~\ref{fig:swomnip}d and Figure~\ref{fig:mshomnip}a,~\ref{fig:mshomnip}b,~\ref{fig:mshomnip}c,~\ref{fig:mshomnip}d. The intervals when the GUMICS$-$4 simulations and the Cluster SC3 measurements disagreed are collected for intervals in the solar wind (Table~\ref{tab:omnisw}) and the magnetosheath (Table~\ref{tab:omnimsh}). The definition of disagreement of the simulations and measurements is quite arbitrary. When the two curves deviate or the correlation function is not symmetric we considered the simulations and the measurements disagreeing. The correlation coeffitiens are also high in these cases however the time shift is large ($\sim$60\,min). The averaged shifted OMNI parameters of the poorly agreeing intervals from the Tables~\ref{tab:omnisw}~and~\ref{tab:omnimsh} are saved. The distributions of the OMNI parameters belonging to the bad simulation results are calculated for the solar wind region (Figure~\ref{fig:swomnibxyz},~\ref{fig:swomnivxyz}~and~\ref{fig:swomnip}) and in the magnetosheath (Figure~\ref{fig:mshomnibxyz},~\ref{fig:mshomnivxyz}~and~\ref{fig:mshomnip}).
\begin{enumerate}
\item In the solar wind the distributions of the OMNI $B_{x}$, $B_{y}$ and $B_{z}$ can be compared in Figure~\ref{fig:swomnibxyz}a,~\ref{fig:swomnibxyz}d,~\ref{fig:swomnibxyz}g,~\ref{fig:swomnibxyz}j; Figure~\ref{fig:swomnibxyz}b,~\ref{fig:swomnibxyz}e,~\ref{fig:swomnibxyz}h,~\ref{fig:swomnibxyz}k; furthermore in Figure~\ref{fig:swomnibxyz}c,~\ref{fig:swomnibxyz}f,~\ref{fig:swomnibxyz}i,~\ref{fig:swomnibxyz}l.
\begin{enumerate}
\item When the $B_{z}$ disagrees in simulations and measurements in Figure~\ref{fig:swomnibxyz}a,~\ref{fig:swomnibxyz}b,~\ref{fig:swomnibxyz}c the black and red distributions of the OMNI $B_{x}$, $B_{y}$ and $B_{z}$ are not similar. The reason for these strange spikes is that there is only one poorly correlated interval for the $B_{z}$ in the solar wind according to Table~\ref{tab:omnisw}.

\item When the $V_{x}$ disagrees in simulations and measurements in Figure~\ref{fig:swomnibxyz}d,~\ref{fig:swomnibxyz}e,~\ref{fig:swomnibxyz}f the black and red distributions of the OMNI $B_{x}$, $B_{y}$ and $B_{z}$ are similar. The distributions do not agree perfectly because in Table~\ref{tab:omnisw} the number of the poorly correlated intervals is only six for the $V_{x}$ component.

\item When the $n_{CIS}$ disagrees in simulations and measurements in Figure~\ref{fig:swomnibxyz}g,~\ref{fig:swomnibxyz}h,~\ref{fig:swomnibxyz}i the black and red distributions of the OMNI $B_{x}$, $B_{y}$ and $B_{z}$ are similar. The distributions do not agree perfectly because in Table~\ref{tab:omnisw} the number of the poorly correlated intervals is only twelve for the $n_{CIS}$.
 
\item When the $n_{EFW}$ disagrees in Figure~\ref{fig:swomnibxyz}j,~\ref{fig:swomnibxyz}k,~\ref{fig:swomnibxyz}l the black and red distributions of the OMNI $B_{x}$, $B_{y}$ and $B_{z}$ are similar. The distributions do not agree perfectly because in Table~\ref{tab:omnisw} the number of the poorly correlated intervals is only nine for $n_{EFW}$.
\end{enumerate}
The values of the OMNI $B_{x}$, $B_{y}$, and $B_{z}$ are not peculiar in the solar wind.

\item In the solar wind the distributions of the OMNI $V_{x}$, $V_{y}$ and $V_{z}$ can be compared in Figure~\ref{fig:swomnivxyz}a,~\ref{fig:swomnivxyz}d,~\ref{fig:swomnivxyz}g,~\ref{fig:swomnivxyz}j; Figure~\ref{fig:swomnivxyz}b,~\ref{fig:swomnivxyz}e,~\ref{fig:swomnivxyz}h,~\ref{fig:swomnivxyz}k; furthermore in Figure~\ref{fig:swomnivxyz}c,~\ref{fig:swomnivxyz}f,~\ref{fig:swomnivxyz}i,~\ref{fig:swomnivxyz}l.
\begin{enumerate} 
\item When the $B_{z}$ disagrees in Figure~\ref{fig:swomnivxyz}a,~\ref{fig:swomnivxyz}b,~\ref{fig:swomnivxyz}c the black and red distributions of the OMNI $V_{x}$, $V_{y}$ and $V_{z}$ are not similar. The reason for these strange spikes is that there is only one poorly correlated interval for the $B_{z}$ in the solar wind according to Table~\ref{tab:omnisw}.

\item When the $V_{x}$ disagrees in simulations and measurements in Figure~\ref{fig:swomnivxyz}d,~\ref{fig:swomnivxyz}e,~\ref{fig:swomnivxyz}f the black and red distributions of the OMNI $V_{x}$, $V_{y}$ and $V_{z}$ are similar. The distributions do not agree perfectly because in Table~\ref{tab:omnisw} the number of the poorly correlated intervals is only six for the $V_{x}$ component.
 
\item When the $n_{CIS}$ disagrees in Figure~\ref{fig:swomnivxyz}g,~\ref{fig:swomnivxyz}h,~\ref{fig:swomnivxyz}i the black and red distributions of the OMNI $V_{x}$, $V_{y}$ and $V_{z}$ are similar. The distributions do not agree perfectly because in Table~\ref{tab:omnisw} the number of the poorly correlated intervals is only twelve for the $n_{CIS}$.

\item When the $n_{EFW}$ disagrees in simulations and measurements in Figure~\ref{fig:swomnivxyz}j,~\ref{fig:swomnivxyz}k,~\ref{fig:swomnivxyz}l the black and red distributions of the OMNI $V_{x}$, $V_{y}$ and $V_{z}$ are similar. The distributions do not agree perfectly because in Table~\ref{tab:omnisw} the number of the poorly correlated intervals is only nine for the $n_{EFW}$.
\end{enumerate}
The values of the OMNI $V_{x}$, $V_{y}$, and $V_{z}$ are not peculiar in the solar wind. 
 
\item In the solar wind the distributions of the solar wind P calculated from OMNI parameters can be compared in Figure~\ref{fig:swomnip}a,~\ref{fig:swomnip}b,~\ref{fig:swomnip}c,~\ref{fig:swomnip}d.
\begin{enumerate}
\item When the $B_{z}$ disagrees in Figure~\ref{fig:swomnip}a the black and red distributions of the OMNI $P$ are not similar. The reason for this strange spike is that there is only one poorly correlated interval for the $B_{z}$ in the solar wind according to Table~\ref{tab:omnisw}.
  
\item When the $V_{x}$ disagrees in simulations and measurements in Figure~\ref{fig:swomnip}b the black and red distributions of the OMNI $P$ are similar. The distributions do not agree perfectly because in Table~\ref{tab:omnisw} the number of the poorly correlated intervals is only six $V_{x}$ component.

\item When the $n_{CIS}$ disagrees in simulations and measurements in Figure~\ref{fig:swomnip}c the black and red distributions of the OMNI $P$ are similar. The distributions do not agree perfectly because in Table~\ref{tab:omnisw} the number of the poorly correlated intervals is only twelve for the $n_{CIS}$.

\item When the $n_{EFW}$ disagrees in simulations and measurements in Figure~\ref{fig:swomnip}d the black and red distributions of the OMNI $P$ are similar. The distributions do not agree perfectly because in Table~\ref{tab:omnisw} the number of the poorly correlated intervals is only nine for the $n_{EFW}$.
\end{enumerate}
The values of the OMNI $P$ are not peculiar in the solar wind. 

\item In the magnetosheath the distributions of the OMNI $B_{x}$, $B_{y}$ and $B_{z}$ can be compared in Figure~\ref{fig:mshomnibxyz}a,~\ref{fig:mshomnibxyz}d,~\ref{fig:mshomnibxyz}g,~\ref{fig:mshomnibxyz}j; Figure~\ref{fig:mshomnibxyz}b,~\ref{fig:mshomnibxyz}e,~\ref{fig:mshomnibxyz}h,~\ref{fig:mshomnibxyz}k; furthermore in Figure~\ref{fig:mshomnibxyz}c,~\ref{fig:mshomnibxyz}f,~\ref{fig:mshomnibxyz}i,~\ref{fig:mshomnibxyz}l. 
\begin{enumerate}
\item When the $B_{z}$ disagrees in simulations and measurements in Figure~\ref{fig:mshomnibxyz}a,~\ref{fig:mshomnibxyz}b,~\ref{fig:mshomnibxyz}c the black and red distributions of the OMNI $B_{x}$, $B_{y}$ and $B_{z}$ are similar.

\item When the $V_{x}$ disagrees in simulations and measurements in Figure~\ref{fig:mshomnibxyz}d,~\ref{fig:mshomnibxyz}e,~\ref{fig:mshomnibxyz}f the black and red distributions of the OMNI $B_{x}$, $B_{y}$ and $B_{z}$ are similar.
 
\item When the $n_{CIS}$ disagrees in simulations and measurements in Figure~\ref{fig:mshomnibxyz}g,~\ref{fig:mshomnibxyz}h,~\ref{fig:mshomnibxyz}i the black and red distributions of the OMNI $B_{x}$, $B_{y}$ and $B_{z}$ are similar.

\item When the $n_{EFW}$ disagrees in Figure~\ref{fig:mshomnibxyz}j,~\ref{fig:mshomnibxyz}k,~\ref{fig:mshomnibxyz}l the black and red distributions of the OMNI $B_{x}$, $B_{y}$ and $B_{z}$ are similar.
\end{enumerate}
The distributions agree quite well because in Table~\ref{tab:omnimsh} the number of the poorly correlated intervals 18, 50, 33 and 30 for the $B_{z}$, the $V_{x}$, the $n_{CIS}$ and $n_{CIS}$ components, respectively. The number of cases is higher and the values of the OMNI $B_{x}$, $B_{y}$ and $B_{z}$ are not peculiar in the magnetosheath.

\item In the magnetosheath the distributions of the OMNI $V_{x}$, $V_{y}$ and $V_{z}$ can be compared in Figure~{\ref{fig:mshomnivxyz}a,~\ref{fig:mshomnivxyz}d,~\ref{fig:mshomnivxyz}g,~\ref{fig:mshomnivxyz}j; Figure~\ref{fig:mshomnivxyz}b,~\ref{fig:mshomnivxyz}e,~\ref{fig:mshomnivxyz}h,~\ref{fig:mshomnivxyz}k; furthermore in Figure~\ref{fig:mshomnivxyz}c,~\ref{fig:mshomnivxyz}f,~\ref{fig:mshomnivxyz}i,~\ref{fig:mshomnivxyz}l.}
\begin{enumerate}
\item When the $B_{z}$ disagrees in Figure~\ref{fig:mshomnivxyz}a,~\ref{fig:mshomnivxyz}b,~\ref{fig:mshomnivxyz}c the black and red distributions of the OMNI $V_{x}$, $V_{y}$ and $V_{z}$ are similar.

\item When the $V_{x}$ disagrees in simulations and measurements in Figure~\ref{fig:mshomnivxyz}d,~\ref{fig:mshomnivxyz}e,~\ref{fig:mshomnivxyz}f the black and red distributions of the OMNI $V_{x}$, $V_{y}$ and $V_{z}$ are similar.

\item When the $n_{CIS}$ disagrees in Figure~\ref{fig:mshomnivxyz}g,~\ref{fig:mshomnivxyz}h,~\ref{fig:mshomnivxyz}i the black and red distributions of the OMNI $V_{x}$, $V_{y}$ and $V_{z}$ are similar.

\item When the $n_{EFW}$ disagrees in simulations and measurements in Figure~\ref{fig:mshomnivxyz}j,~\ref{fig:mshomnivxyz}k,~\ref{fig:mshomnivxyz}l the black and red distributions of the OMNI $V_{x}$, $V_{y}$ and $V_{z}$ are similar.
\end{enumerate}
The distributions agree quite well because in Table~\ref{tab:omnimsh} the number of the poorly correlated intervals 18, 50, 33 and 30 for the $B_{z}$, the $V_{x}$, the $n_{CIS}$ and $n_{CIS}$ components, respectively. The number of cases is higher and the values of the OMNI $V_{x}$, $V_{y}$ and $V_{z}$ are not peculiar in the magnetosheath.

\item In the magnetosheath the distributions of the solar wind P calculated from OMNI parameters can be compared in Figure~\ref{fig:mshomnip}a,~\ref{fig:mshomnip}b,~\ref{fig:mshomnip}c,~\ref{fig:mshomnip}d.
\begin{enumerate}
\item When the $B_{z}$ disagrees in Figure~\ref{fig:mshomnip}a the black and red distributions of the OMNI $P$ are similar.

\item When the $V_{x}$ disagrees in simulations and measurements in Figure~\ref{fig:mshomnip}b the black and red distributions of the OMNI $P$ are similar.

\item When the $n_{CIS}$ disagrees in simulations and measurements in Figure~\ref{fig:mshomnip}c the black and red distributions of the OMNI $P$ are similar.

\item When the $n_{EFW}$ disagrees in simulations and measurements in Figure~\ref{fig:mshomnip}d the black and red distributions of the OMNI $P$ are similar.
\end{enumerate}
The distributions agree quite well because in Table~\ref{tab:omnimsh} the number of the poorly correlated intervals 18, 50, 33 and 30 for the $B_{z}$, the $V_{x}$, the $n_{CIS}$ and $n_{CIS}$ components, respectively. The number of cases is higher and the values of the OMNI $P$ are not peculiar in the magnetosheath.
\end{enumerate}
The inaccuracy of the GUMICS-4 simulations does not depend on the OMNI parameters in the solar wind and magnetosheath regions. The same study does not need to be done for the magnetosphere because the deviation of the measurements and the simulations is so large that it cannot be caused by wrong OMNI solar wind parameters.

The bow shock positions agree in the GUMICS simulations and the Cluster SC3 measurements. However, the magnetopause locations do not match as well as the bow shock in simulations and observations. In simulations the location of the magnetopause is determined from peaks in currents density, particle density gradient, or changes in flow velocity \cite[see references therein]{siscoe01:_magnet_fluop,garcia07:_findin_lyon_fedder_mobar,gordeev13:_verif_gumic_mhd}. Here the previously saved simulation parameters along the virtual Cluster SC3 orbit are analyzed. The $J_{y}$ current density component cannot readily be determined from measurements by a single spacecraft. Therefore, the above--mentioned methods cannot be applied. This discrepancy of the magnetopause location agrees with the results of \citeA{gordeev13:_verif_gumic_mhd} and \citeA{facsko16:_one_earth}. \citeA{gordeev13:_verif_gumic_mhd} compared synthetic GUMICS runs with an empirical formula for the magnetopause locations. \citeA{facsko16:_one_earth} used OMNI solar wind data as input and got the same result as \citeA{gordeev13:_verif_gumic_mhd} and this paper. The neutral sheets are visible in both simulations and observations (Figure~\ref{fig:nsplot},~Table~\ref{tab:ns}). This experience is exceptional because the night side magnetosphere of the GUMICS$-$4 simulations is small and twisted \cite{gordeev13:_verif_gumic_mhd,facsko16:_one_earth}. However, in these cases, the IMF has no large $B_{y}$ component. From \citeA{facsko16:_one_earth} we know that the GUMICS has a normal long tail (or night side magnetosphere) if the $B_{y}$ is small. The code can identify the bow shock transitions. For the magnetopause and the neutral sheet, the results are more complex.

\section{Summary and conclusions}
\label{sec:concl}

Based on the previously created 1-year long GUMICS$-$4 run global MHD simulation results are compared with Cluster SC3 magnetic field, solar wind velocity, and density measurements along the spacecraft orbit. Intervals are selected when the Cluster SC3 and the virtual space probe are situated in the solar wind, magnetosheath, and magnetosphere; firthermore their correlation is calculated. Bow shock, magnetopause, and neutral sheet crossings are selected and their visibility and relative position are compared. We achieved the following results:
\begin{enumerate}
\item In the solar wind the correlation coefficient of the $B_{z}$, the $V_{x}$, the $n_{EFW}$ and the $n_{CIS}$ are larger than 0.8, 0.9, 0.9 and 0.9, respectively. The agreement of the $B_{z}$, the $V_{x}$, and the $n_{EFW}$ is very good, furthermore the agreement of the $n_{CIS}$ is also good. 
\item In the magnetosheath the correlation coefficient of the $B_{z}$, the $V_{x}$, the $n_{EFW}$ and the $n_{CIS}$ are larger than 0.6, 0.9, 0.9 and 0.9, respectively. The agreement of the magnetic field component, the ion plasma moments, and the calculated empirical density is a bit weaker than in the solar wind. The $V_{x}$, the $n_{EFW}$ and the $n_{CIS}$ fits better than the $B_{z}$ component in the magnetosheath. Their agreement is still good. The reason for the deviation is the turbulent behavior of the slowed down and thermalized turbulent solar wind.
\item In neither the dayside nor the nightside magnetosphere can the GUMICS$-$4 provide realistic results. The simulation outputs and the spacecraft measurement disagree in this region. The reason for this deviation must be the missing coupled inner magnetosphere model. The applied tilted dipole approach is not satisfactory in the magnetosphere at all. 
\item Disagreement between GUMICS$-$4 and observations does not seem to be due to any particular upstream solar wind conditions.
\item The position of the bow shock and the neutral sheet agrees well in the simulations and the Cluster SC3 magnetic field, ion plasma moments, and derived electron density measurements in this study. The position of the magnetopause does not fit that well.
\end{enumerate}
The GUMICS$-$4 has scientific and strategic importance for the European Space Weather and Scientific community. This code developed in the Finnish Meteorological Institute is the most developed and tested, widely used tool for modeling the cosmic environment of the Earth in Europe. An inner magnetosphere model should be two--way coupled to the existing configuration of the simulation tool to improve the accuracy of the simulations.

\acknowledgments
The OMNI data were obtained from the GSFC/SPDF OMNIWeb interface at http://omniweb.gsfc.nasa.gov. The authors thank the FGM Team (PI: Chris Carr), the CIS Team (PI: Iannis Dandouras), the WHISPER Team (PI: Jean-Louis Rauch), the EFW Team (PI: Mats Andre), and the Cluster Science Archive for providing FGM magnetic field, CIS HIA ion plasma and WHISPER electron density measurements. Data analysis was partly done with the QSAS science analysis system provided by the United Kingdom Cluster Science Centre (Imperial College London and Queen Mary, University of London) supported by the Science and Technology Facilities Council (STFC). Eija Tanskanen acknowledges financial support from the Academy of Finland for the ReSoLVE Centre of Excellence (project No.~272157). G{\'a}bor Facsk{\'o} thanks Anna-M\'aria V{\'\i{}}gh for improving the English of the paper. The authors thank Pekka Janhunen for developing the GUMICS$-$4 code; furthermore, Minna Palmroth, Andrea Opitz\textbf{, J{\'a}nos Kalm{\'a}r} and Zolt{\'a}n V{\"o}r{\"o}s for the useful discussions. The work of G{\'a}bor Facsk{\'o} was supported by mission science funding for the Van Allen Probes mission, the American Geophysical Union, and the Institute of Earth Physics and Space Science (ELKH EPSS), Sopron, Hungary. This work was partially financed by the National Research, Development and Innovation Office (NKFIH) FK128548 grant. For further use of the year run data, request the authors or use the archive of the Community Coordinated Modelling Center (https://ccmc.gsfc.nasa.gov/publications/posted\_runs/FacskoEA2016.php).

\bibliography{swe-2021SW002807-tx}


\pagebreak

\pagebreak

\pagebreak

\pagebreak

\pagebreak

\pagebreak

\pagebreak

\pagebreak

\pagebreak

\pagebreak

\pagebreak

\pagebreak

\pagebreak

\pagebreak

\pagebreak

\pagebreak

\pagebreak

\pagebreak

\pagebreak

\pagebreak


\pagebreak

\begin{table}[h]
\setlength{\tabcolsep}{3pt}
\caption{The studied solar wind intervals. The correlation coefficients ($C_{B_{z}}$, $C_{V_{x}}$, $C_{n_{CIS}}$, $C_{n_{EFW}}$) and time shift ($\delta t_{V_{x}}$, $\delta t_{n_{CIS}}$, $\delta t_{n_{EFW}}$) in minutes of the magnetic field GSE Z component ($B_z$), solar wind velocity X component ($V_x$), CIS and EFW densities ($n_{CIS}, n_{EFW}$).\label{tab:sw}}
\centering
\begin{tabular}{c||cc|cc|cc|cc}
\hline
Start/End & $C_{B_{z}}$ & $\delta t_{B_{z}}$ & $C_{V_{x}}$ & $\delta t_{V_{x}}$ & $C_{n_{CIS}}$ & $\delta t_{n_{CIS}}$ & $C_{n_{EFW}}$ & $\delta t_{n_{EFW}}$ \\
& & [min] & & [min] & & [min] & & [min] \\
\hline
20020201 20:00/0203 04:00 & 0.97 & 3 & 1.00 & 12 & 0.96 & 3 & 0.98 & 3 \\
20020211 13:00/0212 12:00 & 0.86 & 2 & 1.00 & 0 & 0.99 & 19 & 0.99 & 18 \\
20020218 09:00/0219 02:00 & 0.95 & 1 & 1.00 & -4 & 1.00 & -3 & 0.97 & -2 \\
20020219 06:30/0219 15:00 & 0.96 & 1 & 0.99 & -1 & 0.99 & -60 & 1.00 & 60 \\
20020220 18:30/0222 00:00 & 0.90 & 4 & 1.00 & 4 & 0.93 & -20 & 0.98 & 3 \\
20020318 17:30/0319 02:30 & 0.91 & 2 & 1.00 & 21 & 0.98 & 51 & 0.99 & 6 \\
20020412 20:30/0413 02:00 & 0.91 & 5 & 0.99 & -53 & 0.94 & 60 & 0.98 & 12 \\
20021227 12:00/1228 03:00 & 0.84 & 4 & 1.00 & -2 & 0.99 & -21 & 0.99 & 22 \\
20021229 20:00/1230 16:00 & 0.76 & 1 & 1.00 & 1 & 0.99 & -30 & 0.98 & 43 \\
20030106 06:00/0106 19:00 & 0.82 & 5 & 1.00 & 7 & 0.99 & 3 & 0.95 & -60 \\
20030108 07:00/0109 03:30 & 0.56 & 10 & 1.00 & 41 & 0.99 & 9 & 0.97 & -56 \\
20030113 08:30/0113 18:00 & 0.94 & 3 & 1.00 & 5 & 1.00 & 3 & 0.97 & -1 \\
20030120 07:30/0120 13:00 & 0.86 & 3 & 1.00 & 8 & 1.00 & 4 & 1.00 & -55 \\
20030122 12:00/0123 14:00 & 0.85 & 2 & 1.00 & 3 & 1.00 & 3 & 0.92 & -60 \\
20030124 18:00/0126 00:00 & 0.78 & 3 & 1.00 & 0 & 0.99 & -60 & 0.99 & 60 \\
20030127 16:00/0128 06:00 & 0.89 & -1 & 1.00 & -3 & 0.96 & 1 & 0.89 & 12 \\
20030129 12:00/0130 18:00 & 0.92 & 2 & 1.00 & 4 & 0.95 & -59 & 0.98 & 1 \\
\hline
\end{tabular}
\end{table}

\pagebreak

\begin{center}
\setlength{\tabcolsep}{3pt}
\begin{longtable}{c||cc|cc|cc|cc}
\caption{The studied magnetosheath intervals. The correlation coefficients ($C_{B_{z}}$, $C_{V_{x}}$, $C_{n_{CIS}}$, $C_{n_{EFW}}$) and time shift ($\delta t_{V_{x}}$, $\delta t_{n_{CIS}}$, $\delta t_{n_{EFW}}$) in minutes of the magnetic field GSE Z component ($B_z$), solar wind velocity X component ($V_x$), CIS and EFW densities ($n_{CIS}, n_{EFW}$). In the empty slots the correlation calculation gives invalid result. \label{tab:msh}}\\
\hline
Start/End & $C_{B_{z}}$ & $\delta t_{B_{z}}$ & $C_{V_{x}}$ & $\delta t_{V_{x}}$ & $C_{n_{CIS}}$ & $\delta t_{n_{CIS}}$ & $C_{n_{EFW}}$ & $\delta t_{n_{EFW}}$ \\
& & [min] & & [min] & & [min] & & [min] \\
\hline
\endfirsthead
\multicolumn{9}{c}%
{\tablename\ \thetable\ -- \textit{Continued from previous page}} \\
\hline
Start/End & $C_{B_{z}}$ & $\delta t_{B_{z}}$ & $C_{V_{x}}$ & $\delta t_{V_{x}}$ & $C_{n_{CIS}}$ & $\delta t_{n_{CIS}}$ & $C_{n_{EFW}}$ & $\delta t_{n_{EFW}}$ \\
& & [min] & & [min] & & [min] & & [min] \\
\hline
\endhead
\hline \multicolumn{9}{r}{\textit{Continued on next page}} \\
\endfoot
\hline
\endlastfoot
20020201 13:30/0201 18:30 & 0.92 & 1 & 0.98 & 57 & 0.99 & 60 & 0.98 & 60 \\
20020208 18:15/0209 00:00 & 0.78 & 3 & 0.95 & 60 & 0.98 & -53 & 0.98 & -54 \\
20020211 02:30/0211 09:00 & 0.81 & 0 & 0.99 & -21 & 1.00 & 0 & 0.99 & 0 \\
20020212 16:30/0212 21:00 & 0.86 & 3 & 1.00 & 54 & 0.99 & 30 & 0.99 & 30 \\
20020219 17:30/0219 23:00 & 0.78 & 4 & 0.99 & 37 & 1.00 & 6 & 1.00 & 6 \\
20020222 23:00/0223 06:30 & 0.69 & 1 & 0.97 & -60 & 0.99 & -52 & 0.99 & -48 \\
20020227 16:30/0227 23:15 & 0.53 & 60 & 0.98 & -31 & 1.00 & -38 & 1.00 & -11 \\
20020310 18:30/0311 00:30 & 0.98 & 3 & 0.98 & 20 & 0.99 & 8 & 0.99 & -2 \\
20020311 14:00/0311 19:00 & 0.88 & 5 & 0.97 & 36 & 0.99 & -3 & 0.99 & -40 \\
20020406 19:00/0407 01:15 & 0.79 & 1 & 0.97 & -60 & 0.98 & -56 & 0.98 & -56 \\
20020410 17:30/0410 23:00 & 0.89 & 5 & 0.99 & -52 & 1.00 & 3 & 1.00 & 5 \\
20020411 11:30/0411 16:30 & 0.84 & 3 & 0.99 & 40 & 0.99 & 3 & 0.99 & 3 \\
20020418 18:30/0418 22:45 & 0.93 & 59 & 0.99 & -60 & 0.99 & 60 & 0.98 & 60 \\
20020421 04:30/0421 07:45 & 0.98 & 55 & 1.00 & -60 & 1.00 & -60 & 1.00 & -60 \\
20020422 11:45/0422 15:45 & 0.77 & -5 & 0.98 & -17 & 0.99 & -15 & 0.99 & -16 \\
20020423 08:30/0423 12:30 & 0.94 & 31 & 1.00 & 4 & 0.99 & 16 & 1.00 & 16 \\
20020430 12:30/0430 17:00 & 0.81 & 58 & 0.99 & 23 & 0.99 & -18 & & \\
20020505 07:00/0505 11:15 & 0.83 & 59 & 0.99 & 32 & 0.99 & -60 & & \\
20020506 19:15/0507 00:15 & 0.89 & -28 & 0.99 & -60 & 0.98 & -36 & & \\
20020507 17:30/0507 23:00 & 0.94 & 1 & 0.99 & 47 & 0.99 & -47 & & \\
20020514 22:45/0515 03:00 & 0.82 & 49 & 0.99 & -60 & 0.99 & 32 & 0.99 & -37 \\
20020517 07:00/0517 12:15 & 0.76 & -6 & 1.00 & -5 & 0.99 & -4 & 0.99 & -3 \\
20020518 13:30/0518 19:30 & 0.76 & 1 & 0.99 & 11 & 0.98 & -2 & 0.98 & -2 \\
20020519 20:00/0520 03:30 & 0.98 & 2 & 1.00 & -9 & 0.99 & -4 & 0.99 & -50 \\
20020520 10:45/0520 20:15 & 0.80 & 1 & 0.99 & -3 & 0.95 & -1 & 0.99 & -1 \\
20020522 02:00/0522 08:45 & 0.53 & 52 & 0.99 & 4 & 0.99 & 11 & 0.99 & 22 \\
20020527 02:15/0527 17:15 & 0.80 & -3 & 0.99 & -2 & 0.98 & 0 & 0.99 & 0 \\
20020530 05:00/0530 10:30 & 0.30 & 3 & 1.00 & -23 & 0.99 & 4 & 0.99 & 3 \\
20020601 19:30/0602 01:00 & 0.68 & -2 & 1.00 & 17 & 0.99 & -6 & 0.99 & -7 \\
20020602 21:45/0603 17:45 & 0.65 & -5 & 0.99 & 0 & 0.98 & 3 & 0.99 & 3 \\
20020605 10:30/0606 06:00 & 0.20 & 0 & 0.99 & -7 & 0.98 & 10 & 0.98 & 9 \\
20020607 18:00/0607 22:00 & 0.93 & -35 & 1.00 & -34 & 0.99 & 16 & 0.99 & 15 \\
20020608 01:15/0608 18:15 & 0.54 & -4 & 1.00 & -39 & 0.97 & -6 & 0.97 & -6 \\
20020610 01:30/0610 09:30 & 0.80 & 5 & 1.00 & 8 & 0.99 & 3 & 1.00 & -7 \\
20020610 11:00/0611 01:00 & 0.89 & -4 & 1.00 & -35 & 0.99 & 24 & 0.99 & 7 \\
20020612 18:30/0613 06:15 & 0.45 & -2 & 0.99 & -7 & 0.97 & -3 & 0.97 & -33 \\
20020615 07:00/0615 23:30 & & & 1.00 & 47 & 0.98 & -3 & 0.98 & -5 \\
20020617 05:00/0618 03:45 & 0.79 & 3 & 1.00 & 28 & 0.98 & 9 & 0.99 & 8 \\
20020620 04:00/0620 11:00 & 0.65 & -8 & 0.99 & -6 & 0.98 & 11 & 0.98 & 6 \\
20020622 14:30/0622 18:00 & 0.99 & 56 & 1.00 & 33 & 1.00 & 16 & 1.00 & 16 \\
20021201 04:15/1202 07:45 & 0.41 & 1 & 1.00 & 2 & 0.99 & 6 & 0.99 & 6 \\
20021203 15:30/1204 19:30 & 0.72 & 1 & 0.99 & 60 & 0.98 & 59 & 0.98 & 59 \\
20021207 00:30/1207 07:45 & 0.53 & 38 & 0.99 & -50 & 0.99 & -20 & 0.99 & 20 \\
20021208 09:30/1209 08:00 & 0.72 & 3 & 0.99 & -36 & 0.98 & 5 & 0.98 & 5 \\
20021212 23:30/1213 14:30 & 0.53 & 5 & 1.00 & 36 & 0.99 & -3 & 0.95 & -56 \\
20021213 21:15/1214 09:30 & 0.96 & 5 & 1.00 & -35 & 0.99 & -5 & 0.99 & -46 \\
20021215 12:45/1216 18:00 & 0.80 & 2 & 0.99 & -60 & 0.95 & -60 & 0.98 & 30 \\
20021217 16:30/1218 01:45 & 0.91 & 2 & 1.00 & -54 & 0.99 & 3 & 0.99 & 3 \\
20021220 01:30/1220 06:15 & 0.93 & 0 & 1.00 & 60 & 0.99 & 2 & 0.99 & 3 \\
20021223 02:15/1223 13:00 & 0.93 & 1 & 0.97 & 39 & 0.94 & 50 & 0.99 & -14 \\
20021223 14:00/1223 22:30 & 0.88 & 1 & 1.00 & -2 & 0.99 & -1 & 1.00 & -3 \\
20021224 19:00/1225 01:45 & 0.96 & 0 & 1.00 & -43 & 0.99 & 12 & 0.99 & 28 \\
20021225 23:45/1226 07:15 & 0.97 & 7 & 1.00 & -18 & 0.99 & 56 & 0.99 & 56 \\
20021226 23:00/1227 09:45 & 0.83 & 2 & 1.00 & 2 & 0.99 & 4 & 0.99 & 2 \\
20021229 11:45/1229 17:00 & 0.63 & 2 & 1.00 & -32 & 0.99 & 49 & 0.99 & 48 \\
20021230 17:45/1231 01:00 & 0.74 & 1 & 0.99 & 55 & 0.98 & 60 & 0.98 & 22 \\
20021231 23:00/0101 05:15 & 0.92 & 2 & 1.00 & 0 & 0.99 & -54 & 1.00 & -56 \\
20030105 14:00/0105 21:00 & 0.73 & 1 & 1.00 & 1 & 1.00 & -60 & 0.99 & -60 \\
20030106 23:15/0107 03:00 & 0.70 & 4 & 0.99 & 41 & 1.00 & 56 & 1.00 & -60 \\
20030109 08:45/0109 16:15 & & & 0.91 & -55 & 0.98 & -13 & 0.98 & -25 \\
20030110 07:15/0110 15:15 & 0.95 & 1 & 0.99 & -7 & 0.99 & 2 & 0.98 & 11 \\
20030111 08:15/0111 22:30 & 0.88 & 1 & 0.99 & -59 & 0.94 & -15 & 0.94 & 8 \\
20030112 17:30/0113 00:15 & 0.98 & 0 & 1.00 & -47 & 0.99 & 39 & 0.99 & 51 \\
20030114 00:30/0114 08:30 & 0.86 & -1 & 0.99 & -60 & 0.98 & 23 & 0.98 & 8 \\
20030116 10:15/0116 17:45 & 0.64 & 60 & 0.93 & 52 & 0.99 & 60 & 0.99 & 30 \\
20030117 09:30/0117 13:30 & 0.70 & -3 & 1.00 & 7 & 1.00 & -31 & 1.00 & -33 \\
20030118 23:30/0119 03:45 & 0.97 & 3 & 1.00 & -12 & 1.00 & 7 & 0.99 & 7 \\
20030119 21:00/0120 01:00 & 0.96 & 3 & 1.00 & 6 & 1.00 & 38 & 1.00 & 20 \\
20030121 06:30/0121 11:30 & 0.87 & -3 & 0.98 & 40 & 0.99 & 8 & 1.00 & 8 \\
20030122 04:45/0122 09:30 & 0.76 & -2 & 1.00 & 1 & 1.00 & -7 & 1.00 & -4 \\
20030126 01:45/0126 06:30 & 0.90 & 3 & 0.99 & -15 & 1.00 & -51 & 0.99 & 24 \\
20030127 08:15/0127 13:00 & 1.00 & 10 & 1.00 & -60 & 0.99 & -1 & 0.99 & 1 \\
20030128 12:30/0128 17:15 & 0.77 & 60 & 0.99 & -22 & 0.99 & -5 & 0.99 & 21 \\
20030130 19:45/0131 00:15 & 0.98 & 2 & 0.99 & 52 & 0.99 & 8 & 0.99 & 8 \\
\end{longtable}
\end{center}

\pagebreak

\begin{center}
\setlength{\tabcolsep}{3pt}
\begin{longtable}{c}
\caption{The studied magnetosphere intervals (UT).\label{tab:msph}}\\
\hline
Start/End\\
\hline
\endfirsthead
\multicolumn{1}{c}%
{\tablename\ \thetable\ -- \textit{Continued from previous page}} \\
\hline
Start/End\\
\hline
\endhead
\hline \multicolumn{1}{r}{\textit{Continued on next page}} \\
\endfoot
\hline
\endlastfoot
20020213 23:00/0214 01:30 \\
20020217 18:30/0218 02:00 \\
20020220 00:45/0220 12:00 \\
20020222 11:15/0222 20:15 \\
20020225 02:15/0225 08:30 \\
20020227 06:00/0227 12:00 \\
20020302 00:00/0302 03:15 \\
20020306 10:00/0306 18:30 \\
20020308 17:30/0309 06:00 \\
20020311 02:15/0311 12:00 \\
20020313 11:15/0314 00:15 \\ 
20020316 04:45/0316 08:00 \\
20020318 09:00/0318 14:45 \\ 
20020320 20:30/0320 23:55 \\
20020323 04:00/0323 09:45 \\
20020327 23:45/0328 06:15 \\
20020330 07:15/0330 12:45 \\ 
20020401 19:30/0401 22:00 \\
20020406 09:30/0406 18:00 \\
20020408 15:00/0409 00:00 \\
20020410 23:30/0411 09:45 \\
20020413 08:30/0413 19:00 \\ 
20020416 18:00/0417 04:30 \\
20020418 06:00/0418 12:00 \\
20020420 15:00/0420 23:00 \\
20020422 20:00/0423 07:00 \\
20020425 08:30/0425 18:00 \\
20020430 04:40/0430 12:00 \\
20020504 14:30/0504 16:45 \\
20020505 02:30/0505 07:00 \\
20020507 01:30/0507 15:45 \\
20020508 11:00/0510 04:15 \\
20020512 02:45/0512 09:30 \\
20020514 10:30/0514 12:45 \\
20020519 00:30/0519 19:30 \\
20020521 01:30/0521 22:00 \\
20020523 23:30/0524 02:00 \\
20020524 19:00/0525 08:15 \\
20020526 07:30/0526 10:30 \\
20020528 20:00/0529 05:00 \\
20020531 02:15/0531 13:30 \\
20020602 04:30/0602 07:30 \\
20020602 12:00/0602 21:30 \\
20020604 08:30/0605 07:00 \\
20020606 14:30/0607 16:30 \\
20020609 06:00/0609 20:00 \\
20020611 11:00/0612 13:00 \\
20020614 01:00/0614 16:00 \\
20020616 08:00/0616 18:00 \\
20020620 13:30/0622 01:00 \\
20020623 13:00/0623 17:00 \\
20020624 04:00/0624 10:15 \\
20020630 17:45/0701 15:00 \\
20020701 21:00/0703 10:30 \\
20020703 23:00/0706 03:15 \\
20020707 01:00/0708 23:00 \\
20020710 11:30/0714 03:30 \\
20020714 15:45/0715 15:30 \\
20020716 23:30/0717 16:00 \\
20020718 05:45/0722 11:00 \\
20020722 23:45/0728 01:00 \\ 
20020728 02:00/0804 03:45 \\ 
20020804 04:45/0811 06:15 \\ 
20020811 07:30/0816 01:00 \\ 
20020816 15:30/0818 09:00 \\
20020818 10:00/0825 11:30 \\ 
20020825 13:00/0901 14:15 \\ 
20020901 17:15/0903 23:30 \\ 
20020905 02:15/0906 16:30 \\ 
20020907 10:30/0908 17:00 \\ 
20020908 18:00/0915 19:30 \\ 
20020915 21:00/0922 22:30 \\ 
20020923 00:00/0923 23:30 \\
20020924 03:30/0928 22:45 \\ 
20020928 23:30/0930 01:00 \\
20020930 02:15/1006 17:00 \\ 
20021006 17:45/1007 03:30 \\
20021007 05:00/1007 17:30 \\
20021008 07:30/1010 22:00 \\ 
20021010 22:30/1012 22:30 \\
20021012 23:00/1014 06:30 \\
20021014 09:00/1016 04:00 \\ 
20021016 14:00/1019 00:15 \\ 
20021019 01:30/1019 22:00 \\
20021021 04:00/1022 19:30 \\
20021022 22:30/1026 02:30 \\
20021026 04:00/1029 20:15 \\
20021030 01:30/1102 08:00 \\
20021102 22:00/1104 22:00 \\
20021106 00:00/1107 18:00 \\
20021108 02:00/1109 18:45 \\
20021111 00:00/1112 01:30 \\
20021113 03:45/1114 14:15 \\
20021115 20:30/1116 23:00 \\
20021118 01:00/1118 23:30 \\
20021120 17:00/1121 06:00 \\
20021122 21:30/1124 01:00 \\
20021125 04:00/1126 08:30 \\
20021127 20:00/1128 18:30 \\
20021130 04:00/1201 01:30 \\
20021202 14:30/1203 09:00 \\
20021204 22:00/1205 19:30 \\
20021207 09:00/1207 16:30 \\
20021207 18:00/1207 22:00 \\
20021209 16:30/1210 14:30 \\
20021212 13:45/1212 21:30 \\
20021214 13:30/1214 20:00 \\
20021214 21:00/1215 07:30 \\
20021216 21:00/1217 15:00 \\
20021219 08:00/1219 19:30 \\
20021221 15:45/1221 23:15 \\
20021222 00:30/1222 08:45 \\
20021224 02:30/1224 14:00 \\
20021226 10:00/1226 19:00 \\
20021228 19:30/1229 02:30 \\
20021229 04:00/1229 10:00 \\
20021231 05:00/1231 18:45 \\
20030102 12:30/0102 20:45 \\
20030104 20:45/0105 06:00 \\
20030105 07:00/0105 13:30 \\
20030107 05:45/0107 21:00 \\
20030109 17:00/0110 00:45 \\
20030112 00:00/0112 09:15 \\
20030112 10:30/0112 16:00 \\
20030114 11:00/0114 20:00 \\
20030116 20:30/0116 22:45 \\
20030119 04:30/0119 09:30 \\
20030119 14:00/0119 17:00 \\
20030121 13:30/0121 21:30 \\
20030126 07:30/0126 15:45 \\
20030128 17:45/0129 08:15 \\
20030131 01:30/0131 11:45 \\
\end{longtable}
\end{center}

\pagebreak

\begin{center}
\setlength{\tabcolsep}{3pt}
\begin{longtable}{lc}
\caption{Intervals around the studied bow shock crossings. The Cluster SC3 crossed the bow shock in all cases. The 2nd column shows whether the bow shock is visible in the GUMICS$-$4 simulations. \label{tab:bs}}\\
\hline
Start/End & GUMICS Bow Shock \\
\hline
\endfirsthead
\multicolumn{1}{c}%
{\tablename\ \thetable\ -- \textit{Continued from previous page}} \\
\hline
Start/End & GUMICS Bow Shock \\
\hline
\endhead
\hline \multicolumn{1}{r}{\textit{Continued on next page}} \\
\endfoot
\hline
\endlastfoot
20020201 12:00/0202 00:00 & + \\
20020203 00:00/0203 12:00 & + \\
20020206 06:00/0206 18:00 & + \\
20020208 18:00/0209 06:00 & + \\
20020211 06:00/0211 18:00 & + \\
20020212 12:00/0212 18:00 & + \\
20020213 12:00/0213 18:00 & + \\
20020216 00:00/0216 12:00 & + \\
20020217 06:00/0217 12:00 & -- \\
20020218 06:00/0218 18:00 & + \\
20020219 00:00/0219 18:00 & + \\
20020220 12:00/0221 00:00 & + \\
20020221 18:00/0222 00:00 & + \\
20020301 06:00/0301 12:00 & + \\
20020304 12:00/0304 18:00 & + \\
20020306 00:00/0306 06:00 & + \\
20020307 00:00/0307 06:00 & + \\
20020308 06:00/0308 12:00 & + \\
20020309 06:00/0309 12:00 & + \\
20020310 12:00/0311 00:00 & + \\
20020311 18:00/0312 00:00 & + \\
20020313 00:00/0313 06:00 & -- \\
20020314 00:00/0314 12:00 & + \\
20020316 06:00/0316 18:00 & + \\
20020318 12:00/0319 00:00 & + \\
20020323 12:00/0323 18:00 & + \\
20020325 18:00/0326 06:00 & -- \\
20020327 06:00/0327 12:00 & + \\
20020329 18:00/0330 00:00 & -- \\
20020402 00:00/0402 06:00 & + \\
20020405 18:00/0406 00:00 & -- \\
20020407 00:00/0407 06:00 & -- \\
20020409 06:00/0409 12:00 & -- \\
20020410 12:00/0410 18:00 & -- \\
20020411 12:00/0411 18:00 & -- \\
20020413 00:00/0413 06:00 & + \\
20020413 18:00/0414 06:00 & + \\
20020420 00:00/0420 06:00 & + \\
20020423 12:00/0423 23:00 & + \\
20020427 00:00/0427 06:00 & + \\
20020428 06:00/0428 12:00 & + \\
20020430 18:00/0501 00:00 & + \\
20020505 06:00/0505 18:00 & -- \\
20020507 18:00/0509 06:00 & + \\
20020510 06:00/0510 12:00 & + \\
20020513 12:00/0513 18:00 & + \\
20020515 00:00/0515 06:00 & -- \\
20020520 00:00/0520 06:00 & + \\
20020522 06:00/0522 12:00 & + \\
20020522 18:00/0523 06:00 & + \\
20021206 06:00/1207 06:00 & + \\
20021218 00:00/1219 00:00 & + \\
20021220 18:00/1221 00:00 & + \\
20021221 00:00/1221 12:00 & + \\
20021222 12:00/1223 00:00 & + \\
20021223 00:00/1223 06:00 & + \\
20021225 06:00/1226 00:00 & + \\
20021227 06:00/1228 00:00 & + \\
20021228 00:00/1228 12:00 & + \\
20021229 12:00/1230 00:00 & + \\
20030101 06:00/0102 00:00 & + \\
20030103 06:00/0103 12:00 & + \\
20030104 00:00/0104 18:00 & + \\
20030106 00:00/0107 00:00 & + \\
20030108 00:00/0108 12:00 & + \\
20030113 00:00/0114 06:00 & + \\
20030115 00:00/0115 12:00 & + \\
20030118 18:00/0119 00:00 & + \\
20030120 00:00/0121 12:00 & + \\
20030122 06:00/0122 12:00 & + \\
20030123 12:00/0124 00:00 & + \\
20030124 12:00/0124 18:00 & + \\
20030126 00:00/0126 06:00 & + \\
20030127 00:00/0127 18:00 & + \\
20030128 06:00/0128 18:00 & + \\
20030129 06:00/0129 12:00 & + \\
20030130 18:00/0131 00:00 & + \\
\hline
\end{longtable}
\end{center}

\pagebreak

\begin{center}
\setlength{\tabcolsep}{3pt}
\begin{longtable}{lc}
\caption{Intervals around the studied magnetopause crossings. The Cluster SC3 crossed the magnetopause in all cases. The 2nd column shows whether the magnetopause is visible in the GUMICS$-$4 simulations. \label{tab:mp}}\\
\hline
Start/End & GUMICS Magnetopause \\
\hline
\endfirsthead
\multicolumn{1}{c}%
{\tablename\ \thetable\ -- \textit{Continued from previous page}} \\
\hline
Start/End & GUMICS Magnetopause \\
\hline
\endhead
\hline \multicolumn{1}{r}{\textit{Continued on next page}} \\
\endfoot
\hline
\endlastfoot
20020203 06:00/0203 12:00 & + \\
20020206 06:00/0206 12:00 & -- \\
20020211 00:00/0211 06:00 & + \\
20020218 00:00/0218 06:00 & + \\
20020225 06:00/0225 12:00 & + \\
20020302 00:00/0302 06:00 & + \\
20020306 18:00/0307 00:00 & -- \\
20020308 12:00/0308 18:00 & -- \\
20020311 12:00/0311 18:00 & + \\
20020313 18:00/0314 00:00 & -- \\
20020314 00:00/0314 06:00 & + \\
20020323 06:00/0323 12:00 & + \\
20020330 12:00/0330 18:00 & -- \\
20020404 06:00/0404 12:00 & -- \\
20020409 00:00/0409 06:00 & -- \\
20020418 12:00/0418 18:00 & + \\
20020422 12:00/0422 18:00 & -- \\
20020429 18:00/0430 00:00 & -- \\
20020507 12:00/0507 18:00 & -- \\
20020509 06:00/0509 12:00 & -- \\
20020510 00:00/0510 06:00 & -- \\
20020514 18:00/0515 00:00 & -- \\
20020519 12:00/0519 18:00 & -- \\
20020520 12:00/0521 00:00 & -- \\
20020522 00:00/0522 06:00 & -- \\
20020529 00:00/0529 12:00 & -- \\
20020530 06:00/0530 18:00 & -- \\
20020531 18:00/0601 00:00 & -- \\
20020602 18:00/0603 00:00 & -- \\
20020604 06:00/0604 12:00 & -- \\
20020605 06:00/0606 18:00 & -- \\
20020607 12:00/0608 06:00 & + \\
20020609 00:00/0609 06:00 & -- \\
20020610 00:00/0610 06:00 & -- \\
20020611 00:00/0611 12:00 & -- \\
20020612 06:00/0614 00:00 & -- \\
20020614 18:00/0615 06:00 & -- \\
20020616 00:00/0616 12:00 & + \\
20020620 00:00/0620 18:00 & -- \\
20020622 06:00/0622 18:00 & -- \\
20020704 12:00/0705 00:00 & -- \\
20020706 00:00/0706 12:00 & + \\
20020709 00:00/0709 18:00 & -- \\
20020715 18:00/0716 12:00 & -- \\
20030105 06:00/0105 18:00 & + \\
20030110 00:00/0110 12:00 & + \\
20030112 12:00/0112 18:00 & -- \\
20030117 06:00/0117 12:00 & + \\
20030121 06:00/0121 12:00 & + \\
20030122 00:00/0122 06:00 & -- \\
20030126 18:00/0127 00:00 & + \\
20030128 12:00/0128 18:00 & + \\
20030129 00:00/0129 12:00 & + \\
20030131 12:00/0201 00:00 & + \\
\hline
\end{longtable}
\end{center}

\pagebreak

\begin{table}[h]
\setlength{\tabcolsep}{3pt}
\centering
\begin{tabular}{lc}
\hline
Start/End & GUMICS Neutral Sheet\\
\hline
20020901 19:00/0902 00:00 & -- \\
20020906 14:00/0906 16:30 & + \\
20020913 17:30/0913 20:00 & + \\
20020918 13:00/0918 14:30 & -- \\
20020920 20:30/0921 02:00 & + \\
20020928 03:00/0928 07:00 & + \\
20021002 16:00/1003 00:00 & -- \\
20021014 12:30/1014 23:00 & + \\
20021017 03:00/1017 04:00 & -- \\
\hline
\end{tabular}
\caption{Intervals around the studied neutral sheet crossings in the tail. The Cluster SC3 crossed the neutral sheet in all cases. The 2nd column shows whether the neutral sheet is visible in the GUMICS$-$4 simulations. \label{tab:ns}}
\end{table}

\pagebreak

\begin{table}[!h]
\centering
\begin{tabular}{l|ccc|cccc}
\hline
& \multicolumn{3}{|c|}{OMNI} & \multicolumn{4}{|c}{Cluster SC3} \\ 
Start/End & $B_{z}$ & $V_{x}$ & P & $B_{z}$ & $V_{x}$ & $n_{CIS}$ & $n_{EFW}$ \\
& [nT] & [km/s] & [$cm^{-3}$] \\
 \hline
20020201 20:00/0203 04:00 & -1.25 & -373.52 & 4.08 & y & y & n & y \\
20020211 13:00/0212 12:00 & 0.03 & -533.11 & 2.18 & y & y & y & y \\
20020218 09:00/0219 02:00 & 2.56 & -362.41 & 3.46 & y & n & n & y \\
20020219 06:30/0219 15:00 & 3.55 & -401.63 & 1.25 & y & y & n & n \\
20020220 18:30/0222 00:00 & 1.95 & -440.18 & 1.96 & y & y & n & y \\
20020318 17:30/0319 02:30 & 3.79 & -429.30 & 15.34 & y & n & n & n \\
20020412 20:30/0413 02:00 & -1.81 & -420.35 & 3.24 & y & n & n & y \\
20021227 12:00/1228 03:00 & 0.09 & -714.40 & 2.72 & y & n & n & y \\
20021229 20:00/1230 16:00 & -0.37 & -526.40 & 2.26 & y & y & n & n \\
20030106 06:00/0106 19:00 & 2.25 & -399.91 & 1.50 & y & n & n & n \\
20030108 07:00/0109 03:30 & -0.58 & -280.80 & 2.97 & n & n & y & n \\
20030113 08:30/0113 18:00 & 0.68 & -397.83 & 1.72 & y & y & y & n \\
20030120 07:30/0120 13:00 & 2.16 & -630.69 & 2.43 & y & y & y & y \\
20030122 12:00/0123 14:00 & 0.13 & -608.96 & 3.41 & y & y & y & n \\
20030124 18:00/0126 00:00 & -0.71 & -739.68 & 2.87 & y & y & n & n \\
20030127 16:00/0128 06:00 & -0.92 & -451.84 & 3.12 & y & y & n & n \\
20030129 12:00/0130 18:00 & -3.09 & -450.00 & 3.96 & y & y & n & y \\
\hline
\end{tabular}
\caption{The average OMNI input parameters in the solar wind and the good/bad agreement of the GUMICS$-$4 simulations to the Cluster $B_{z}$ magnetic field component, the $V_{x}$ solar wind speed component, the $n_{CIS}$ solar wind density measured by the Cluster CIS HIA instrument and the $n_{EFW}$ solar wind density calculated from the spacecraft potential measured by the Cluster EFW instrument in the solar wind.\label{tab:omnisw}}
\end{table}

\pagebreak

\begin{center}
\setlength{\tabcolsep}{3pt}
\begin{longtable}{l|rcc|cccc}
\caption{The average OMNI input parameters in the solar wind and the good/bad agreement of the GUMICS$-$4 simulations to the Cluster $B_{z}$ magnetic field component, the $V_{x}$ solar wind speed component, the $n_{CIS}$ solar wind density measured by the Cluster CIS HIA instrument and the $n_{EFW}$ solar wind density calculated from the spacecraft potential measured by the Cluster EFW instrument in the magnetosheath.\label{tab:omnimsh}}\\
\hline
& \multicolumn{3}{|c|}{OMNI} & \multicolumn{4}{|c}{Cluster SC3} \\ 
Start/End & $B_{z}$ & $V_{x}$ & P & $B_{z}$ & $V_{x}$ & $n_{CIS}$ & $n_{EFW}$ \\
& [nT] & [km/s] & [$cm^{-3}$] \\
\hline
\endfirsthead
\multicolumn{1}{c}%
{\tablename\ \thetable\ -- \textit{Continued from previous page}} \\
\hline
& \multicolumn{3}{|c|}{OMNI} & \multicolumn{4}{|c}{Cluster SC3} \\ 
Start/End & $B_{z}$ & $V_{x}$ & P & $B_{z}$ & $V_{x}$ & $n_{CIS}$ & $n_{EFW}$ \\
& [nT] & [km/s] & [$cm^{-3}$] \\
\hline
\endhead
\hline \multicolumn{1}{r}{\textit{Continued on next page}} \\
\endfoot
\hline
\endlastfoot
20020201 13:30/0201 18:30 & 0.19 & -342.87 & 4.62 & y & n & n & n \\
20020208 18:15/0209 00:00 & -0.48 & -508.16 & 1.61 & y & n & n & n \\
20020211 02:30/0211 09:00 & -1.85 & -425.67 & 1.78 & y & y & y & y \\
20020212 16:30/0212 21:00 & 2.98 & -509.22 & 2.34 & y & n & n & n \\
20020219 17:30/0219 23:00 & 1.46 & -431.50 & 1.46 & y & y & y & y \\
20020222 23:00/0223 06:30 & 0.86 & -391.22 & 1.14 & y & n & n & n \\
20020227 16:30/0227 23:15 & 1.89 & -343.13 & 1.52 & n & n & n & n \\
20020310 18:30/0311 00:30 & -2.81 & -379.46 & 1.78 & y & y & y & y \\
20020311 14:00/0311 19:00 & 1.63 & -371.43 & 2.68 & n & n & n & n \\
20020406 19:00/0407 01:15 & -2.71 & -333.13 & 0.93 & y & n & n & n \\
20020410 17:30/0410 23:00 & 0.31 & -312.43 & 4.42 & n & n & y & y \\
20020411 11:30/0411 16:30 & -1.50 & -494.02 & 4.25 & y & y & n & n \\
20020418 18:30/0418 22:45 & -0.92 & -450.82 & 0.30 & n & n & n & n \\
20020421 04:30/0421 07:45 & 0.40 & -455.69 & 1.37 & n & n & n & n \\
20020422 11:45/0422 15:45 & 0.25 & -419.98 & 1.14 & n & n & y & y \\
20020423 08:30/0423 12:30 & 2.77 & -507.99 & 6.82 & n & n & n & n \\
20020430 12:30/0430 17:00 & 2.15 & -479.51 & 3.02 & n & n & n & n \\
20020505 07:00/0505 11:15 & 0.20 & -336.81 & 1.74 & n & n & n & n \\
20020506 19:15/0507 00:15 & 0.78 & -390.00 & 2.46 & y & n & n & n \\
20020507 17:30/0507 23:00 & 2.87 & -392.40 & 3.49 & y & n & n & n \\
20020514 22:45/0515 03:00 & -2.42 & -414.01 & 1.82 & n & n & n & n \\
20020517 07:00/0517 12:15 & -0.39 & -379.32 & 1.52 & y & y & y & y \\
20020518 13:30/0518 19:30 & 0.63 & -345.87 & 1.59 & n & n & y & y \\
20020519 20:00/0520 03:30 & 4.75 & -408.56 & 1.12 & y & y & y & y \\
20020520 10:45/0520 20:15 & 0.74 & -448.89 & 1.93 & y & y & y & y \\
20020522 02:00/0522 08:45 & -1.07 & -398.12 & 1.63 & n & y & y & y \\
20020527 02:15/0527 17:15 & -3.11 & -542.53 & 2.07 & y & y & y & y \\
20020530 05:00/0530 10:30 & 0.03 & -493.86 & 2.08 & y & n & y & y \\
20020601 19:30/0602 01:00 & -3.38 & -342.27 & 4.16 & y & y & y & y \\
20020602 21:45/0603 17:45 & 0.38 & -435.47 & 1.89 & y & y & y & y \\
20020605 10:30/0606 06:00 & -0.42 & -394.49 & 1.08 & y & y & n & n \\
20020607 18:00/0607 22:00 & -1.60 & -291.85 & 1.80 & y & y & y & y \\
20020608 01:15/0608 18:15 & 0.06 & -335.39 & 2.74 & y & n & y & y \\
20020610 01:30/0610 09:30 & 1.60 & -465.52 & 3.00 & y & y & y & y \\
20020610 11:00/0611 01:00 & -2.27 & -419.86 & 2.16 & y & n & y & y \\
20020612 18:30/0613 06:15 & -1.13 & -351.03 & 1.16 & y & y & y & y \\
20020615 07:00/0615 23:30 & -1.16 & -334.27 & 2.84 & n & n & y & y \\
20020617 05:00/0618 03:45 & 0.78 & -351.47 & 1.87 & y & n & y & y \\
20020620 04:00/0620 11:00 & 0.46 & -485.48 & 1.73 & y & y & y & y \\
20020622 14:30/0622 18:00 & -0.72 & -429.02 & 1.93 & n & n & y & y \\
20021201 04:15/1202 07:45 & -1.09 & -499.23 & 2.62 & y & y & y & y \\
20021203 15:30/1204 19:30 & 0.34 & -449.09 & 2.06 & y & n & n & n \\
20021207 00:30/1207 07:45 & 0.80 & -451.80 & 7.33 & n & n & y & y \\
20021208 09:30/1209 08:00 & 0.60 & -600.27 & 1.49 & y & n & y & y \\
20021212 23:30/1213 14:30 & 0.10 & -337.77 & 1.32 & y & n & n & n \\
20021213 21:15/1214 09:30 & -0.74 & -361.19 & 2.99 & y & n & y & y \\
20021215 12:45/1216 18:00 & 1.32 & -479.48 & 1.53 & y & n & n & n \\
20021217 16:30/1218 01:45 & 4.56 & -393.99 & 2.49 & y & n & y & y \\
20021220 01:30/1220 06:15 & -1.21 & -530.62 & 3.01 & y & n & y & y \\
20021223 02:15/1223 13:00 & -2.32 & -516.12 & 2.22 & y & n & n & n \\
20021223 14:00/1223 22:30 & 0.89 & -519.77 & 2.55 & y & y & y & y \\
20021224 19:00/1225 01:45 & 0.88 & -523.86 & 3.41 & y & n & y & y \\
20021225 23:45/1226 07:15 & -0.61 & -414.38 & 2.21 & y & y & n & n \\
20021226 23:00/1227 09:45 & -1.79 & -618.14 & 6.20 & y & y & y & y \\
20021229 11:45/1229 17:00 & -0.41 & -580.12 & 2.39 & y & n & n & n \\
20021230 17:45/1231 01:00 & -1.01 & -483.60 & 1.93 & y & n & n & y \\
20021231 23:00/0101 05:15 & 0.60 & -418.95 & 1.94 & y & n & n & n \\
20030105 14:00/0105 21:00 & -0.03 & -414.46 & 1.69 & y & n & n & n \\
20030106 23:15/0107 03:00 & -1.62 & -392.29 & 1.56 & n & n & n & n \\
20030109 08:45/0109 16:15 & 1.45 & -272.82 & 2.31 & n & n & n & n \\
20030110 07:15/0110 15:15 & -2.11 & -401.03 & 2.72 & y & n & y & y \\
20030111 08:15/0111 22:30 & -0.20 & -433.33 & 1.24 & y & n & n & y \\
20030112 17:30/0113 00:15 & 1.53 & -389.62 & 1.45 & y & n & n & n \\
20030114 00:30/0114 08:30 & -1.67 & -388.53 & 2.27 & y & n & n & y \\
20030116 10:15/0116 17:45 & -1.20 & -328.91 & 1.22 & n & n & n & n \\
20030117 09:30/0117 13:30 & -1.36 & -327.09 & 2.55 & y & y & y & y \\
20030118 23:30/0119 03:45 & 6.41 & -459.46 & 4.82 & y & y & y & y \\
20030119 21:00/0120 01:00 & 1.52 & -597.95 & 2.38 & y & n & y & y \\
20030121 06:30/0121 11:30 & -1.77 & -670.25 & 1.50 & y & n & n & n \\
20030122 04:45/0122 09:30 & 0.11 & -588.87 & 2.30 & y & n & y & y \\
20030126 01:45/0126 06:30 & -0.24 & -713.82 & 2.75 & y & y & y & y \\
20030127 08:15/0127 13:00 & 7.94 & -509.30 & 0.47 & y & n & y & y \\
20030128 12:30/0128 17:15 & 4.95 & -443.83 & 4.15 & y & y & y & y \\
20030130 19:45/0131 00:15 & 4.21 & -510.33 & 2.63 & y & n & y & y \\
\hline
\end{longtable}
\end{center}

\end{document}